\documentclass[12pt, a4paper]{article}
\pdfoutput=1

\usepackage{amsthm}
\usepackage{amsmath}
\usepackage{natbib}
\usepackage{graphicx}
\usepackage{comment}

\usepackage{amssymb, tikz}
\usepackage[position=bottom]{subfig}

\usetikzlibrary{calc,fit,positioning,shapes,arrows}
\tikzset{>={stealth}}

\bibpunct[, ]{(}{)}{;}{a}{,}{,}

\def\ci{\mathbin{\perp\!\!\!\perp}}
\def\notci{\mathbin{\perp\!\!\!\perp\!\!\!\!\!\!\!\!\!\!\;\diagup\;}}

\newcommand{\poe}{PoE}

\newcommand{\cpar}{\phi}
\newcommand{\stage}{\ell}

\newcommand{\pcomb}{\p_{\mathrm{comb}}}
\newcommand{\pmeld}{\p_{\mathrm{meld}}}

\newcommand{\prepl}{\p_{\mathrm{repl}}}
\newcommand{\preplm}{\p_{\mathrm{repl},m}}
\newcommand{\pnew}{\pagg}

\newcommand{\rnew}{{\mathrm{pool}}}
\newcommand{\fnew}{F_\rnew}
\newcommand{\ralt}{{\mathrm{alt}}}
\newcommand{\palt}{p_\ralt}

\newcommand{\rtransf}{{\mathrm{trans}}}

\newcommand{\ptmeld}{p_\mathrm{meldtrans}}

\newcommand{\pagg}{\p_{\mathrm{pool}}}
\newcommand{\aggpar}{w}

\newcommand{\mhprob}{r}

\newcommand{\prop}{\star}
\newcommand{\pprop}{q}
\newcommand{\data}{Y}
\newcommand{\obsdata}{y}
\newcommand{\latent}{\psi}

\newcommand{\normalising}{K}

\newcommand{\dunif}{\text{Unif}}

\newcommand{\dinvgamma}{\text{Inv-Gam}}
\newcommand{\dmult}{\text{Mult}}
\newcommand{\dnorm}{\text{N}}
\newcommand{\dlnorm}{\text{Lognormal}}
\newcommand{\dbin}{\text{Bin}}
\newcommand{\dpoi}{\text{Po}}
\newcommand{\dbeta}{\text{Beta}}

\newcommand{\logit}{\text{logit}}

\newcommand{\distributedas}{\sim}
\newcommand{\given}{\mid}
\newcommand{\modelIndex}{m}
\newcommand{\modelFirst}{1}
\newcommand{\modelSecond}{2}
\newcommand{\modelLast}{M}

\newcommand{\poolfun}{g}

\newcommand{\p}{p}

\newcommand{\pone}{\p_{1}}
\newcommand{\ptwo}{\p_{2}}

\newcommand{\mcFirst}{1}
\newcommand{\mcLast}{H}
\newcommand{\mcIndex}{h}
\newcommand{\mcIndexRandom}{d}

\newcommand{\talltheta}{\cpar}
\newcommand{\talldata}{y}
\newcommand{\batchIndex}{b}
\newcommand{\batchFirst}{1}
\newcommand{\batchLast}{B}
\newcommand{\batchSize}{\batchLast}

\newcommand{\flutime}{t}
\newcommand{\flutimetwo}{v}
\newcommand{\flutimetwoset}{V}
\newcommand{\flutimeint}{u}
\newcommand{\flustartofweekset}{U}
\newcommand{\flutimeset}{T}
\newcommand{\flustartofweek}{U}
\newcommand{\fluage}{a}

\newcommand{\flupos}{\pi^{\text{pos}}}
\newcommand{\fluposmin}{\omega}
\newcommand{\fluicudata}{\obsdata}
\newcommand{\fluypos}{z^\text{pos}}
\newcommand{\flunpos}{n^\text{pos}}
\newcommand{\flulambda}{\lambda}
\newcommand{\flugamma}{\gamma}
\newcommand{\flumean}{\eta}
\newcommand{\fluexitrate}{\mu}
\newcommand{\fluexitratealpha}{\alpha}
\newcommand{\fluexitratebeta}{\beta}

\newcommand{\fluna}{\chi}
\newcommand{\fludetprob}{\pi^{\text{det}}}
\newcommand{\fluicugenericpara}{\theta}

\newcommand{\fluicugenericposdata}{\flupos}

\newcommand{\indexdata}{x}
\newcommand{\brookst}{t}

\newcommand{\brookschild}{C}
\newcommand{\brooksadult}{A}
\newcommand{\recoverydata}{y}

\newcommand{\obsprob}{\pi}
\newcommand{\recoveryrate}{\lambda}
\newcommand{\survivalrate}{\eta}
\newcommand{\brooksyearindex}{u}
\newcommand{\truecounts}{\mu}

\newcommand{\indexsd}{\sigma}
\newcommand{\productivityrate}{\gamma}
\newcommand{\brooksalpha}{\alpha}
\newcommand{\brooksbeta}{\beta}
\newcommand{\brooksalphaadult}{\cpar_{\brooksalpha, \brooksadult}}
\newcommand{\brooksalphachild}{\cpar_{\brooksalpha, \brookschild}}
\newcommand{\brooksbetaadult}{\cpar_{\brooksbeta, \brooksadult}}
\newcommand{\brooksbetachild}{\cpar_{\brooksbeta, \brookschild}}
\newcommand{\freezing}{z}
\newcommand{\brooksagegroupindex}{G}
\newcommand{\brooksalphaagegroup}{\cpar_{\brooksalpha, \brooksagegroupindex}}
\newcommand{\brooksbetaagegroup}{\cpar_{\brooksbeta, \brooksagegroupindex}}
\newcommand{\brooksrecoverypara}{\Omega_{1}}

\newcommand{\brookssharedpara}{\Omega_{0}}

\newcommand{\psiall}{\psi_1,\ldots,\allowbreak\psi_M}
\newcommand{\psiallpart}{\psi_1,\ldots,\allowbreak\psi_{\stage-1}}
\newcommand{\psiallstar}{\psi_1^\star,\ldots,\allowbreak\psi_M^\star}
\newcommand{\psiallstarpart}{\psi_1^\star,\ldots,\allowbreak\psi_{\stage-1}^\star}
\newcommand{\psionestar}{\psi_1,\ldots,\allowbreak\psi_m^\star,\ldots,\allowbreak\psi_M}
\newcommand{\yall}{y_1,\ldots,\allowbreak y_M}
\newcommand{\yallpart}{y_1,\ldots,\allowbreak y_{\stage-1}}

\newcommand{\transfun}{Q}

\begin{document}

\title{Joining and splitting models with Markov melding}

\author{
Robert J. B. Goudie, Anne M. Presanis, David Lunn,\\
Daniela De Angelis and Lorenz Wernisch
}

\maketitle
\begin{center}
\emph{MRC Biostatistics Unit, University of Cambridge, UK}%
\end{center}

\begin{abstract}\noindent
Analysing multiple evidence sources is often feasible only via a modular approach, with separate submodels specified for smaller components of the available evidence.
Here we introduce a generic framework that enables fully Bayesian analysis in this setting.
We propose a generic method for forming a suitable joint model when {\em joining\/} submodels, and a convenient computational algorithm for fitting this joint model in stages, rather than as a single, monolithic model.
The approach also enables {\em splitting\/} of large joint models into smaller submodels, allowing inference for the original joint model to be conducted via our multi-stage algorithm.
We motivate and demonstrate our approach through two examples: joining components of an evidence synthesis of A/H1N1 influenza, and splitting a large ecology model.

\end{abstract}

\section{Introduction}
The increasing availability of large amounts of diverse types of data in all scientific fields has prompted an explosion in applications of methods that combine multiple sources of evidence using (Bayesian) graphical models \citep[for example,][]{MoranClark2011,Commenges:2012gv,Shubin2016,Birrell2016}.
Such evidence synthesis methods have several advantages \citep{Ades:2006tw,Welton:2012,Jackson2015}: resulting estimates are typically more precise, due to the increased amount of information;  they are consistent with all available knowledge; and the risk of potential biases introduced if estimation relies on a `best quality' subset is minimised.

However, dealing with joint models of several sources of evidence, including data and expert opinion, may be inferentially imprudent, computationally challenging, or even infeasible. It is often sensible to take a modular approach, where separate submodels are specified for smaller components of the available data, facilitating computation and, importantly, allowing insight into the influence of each submodel on the joint model inference
\citep{hssIntroduction2003, Liu:2009hm}. 
These submodels can originate in two ways: either by first specifying submodels that, in a Bayesian framework, should be \emph{joined} in a single model to allow all information and uncertainty to be fully propagated; or as a result of \emph{splitting} an existing joint model.

\begin{figure}[t]
\centering

\begin{tikzpicture}[minimum width=0.75cm, minimum height = 0.75cm]
\node[circle, draw] (link) at (0, 1.5) {\(\cpar\)};

\node[ellipse, draw] (model1) at (-2, 0) {\(\psi_{1}, \data_{1}\)};
\node[ellipse, draw] (modelM) at (2, 0) {\(\psi_{\modelLast}, \data_{\modelLast}\)};

\node (dots) at (0, 0) {\dots};

\draw[->] (link) -- (model1);
\draw[->] (link) -- (modelM);

\end{tikzpicture}
\caption{DAG representation of a joint hierarchical model linking \(\modelLast\) submodels.}
\label{fig:markov-combination}
\end{figure}
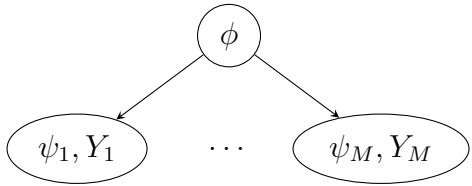

Formally, consider $M$ probability submodels $p_m(\phi,\psi_m,Y_m)$, $m=1,\ldots,M$, for sub\-model-specific multivariate parameters $\psi_m$ and observable random variables $Y_m$, as well as a multivariate parameter $\phi$ common to all submodels that acts as a `link' between the submodels.
The problem is then to {\em join\/} the submodels into a single model \(\pcomb(\phi, \psi_1, \ldots ,\psi_M, \allowbreak Y_1, \ldots ,Y_M)\) so that the posterior distributions for the link parameter \(\cpar\) and the submodel-specific parameters \(\psi_{\modelIndex}\) account for all observations and uncertainty.
A suitable joint model for a collection of submodels naturally arises in some contexts from standard model constructs, such as a hierarchical model (Figure~\ref{fig:markov-combination}).
However, it is not immediately clear how to form such a joint model when either: the submodels are not expressed in a form conditional upon the link parameter \(\cpar\), particularly if the link parameter is a non-invertible deterministic function of the other parameters; or the prior marginal distributions \(p_m(\cpar)\), \(m=1, \ldots, M\), for the link parameter \(\cpar\) differ in the submodels.
In applied research, convenient approximate two-stage approaches have
been widely used, where one submodel is fitted and an approximation of the resulting posterior is provided to a second submodel \citep{Jackson:2009km, Presanis:2014em}.
However, the joint model that is implied by such an approach is unclear \citep{EddyEtAl1992, Ades:2006tw}.

Conversely, suppose a joint model \(\p(\phi,\psi_1,\ldots,\psi_M,Y_1,\ldots,Y_M)\) exists that needs {\em splitting\/}  into $M$ submodels $p_m(\phi,\psi_m,Y_m)$, $m=1,\ldots,M$.
The submodels should be faithful to the original model in the sense that joining the submodels results in the original model.
In some contexts, suitable submodels arise naturally from the
structure of the joint model, resulting in splitting
strategies used implicitly in the context of hierarchical models \citep{Lunn:2013uo, Tom:2010uf, Liang:2007fi} and of tall data \citep{Scott:2013wza, Neiswanger:2013ug}.
However, neither the general conditions stipulating when splitting is permissible nor a general framework for splitting a model are immediately clear.

In this paper we introduce {\em Markov melding\/}, a simple, generic approach for joining and splitting models that 
clarifies and generalises various proposals made in the literature under the umbrella of one theoretical framework.
Markov melding builds on the theory of {\em Markov combination\/} \citep{Dawid:1993fc} and \emph{super Markov combination} \citep{Massa:2010uf, Massa:2017bz} and combines it with ideas from {\em Bayesian melding\/} \citep{Poole:2000ee}, enabling evidence
synthesis \citep{Ades:2006tw, Welton:2012} and model expansion \citep{Draper:1995ts} in realistic applied settings.
Markov combination is a framework for combining submodels when the prior marginal distributions \(p_m(\cpar)\), \(m = 1, \dots, M\), are identical.
Our approach relaxes this assumption, which is often not satisfied in applied settings, to allow joining submodels with similar but not identical prior marginal distributions.
It also accounts for contexts where the link parameter \(\cpar\) is a non-invertible deterministic function of other parameters in a submodel.
When joining submodels, Markov melding aims to preserve the original submodels as faithfully as possible, and, in particular, always preserves the submodel-specific conditional distributions \(p_m(\psi_m, Y_m \mid \phi)\) for all \(\modelIndex\).
Note that, while Markov melding is defined for any collection of submodels, the results may be misleading if any evidence components (priors, submodels and data) strongly conflict \citep{Presanis:2013hb,Gasemyr:2009jt}.
Such conflict should be investigated and resolved, for example through bias modelling \citep{Turner:2009}, before proceeding with the synthesis.
In terms of splitting, the Markov melding framework proposed here clarifies the conditions required and the general framework in which to conduct model splitting, facilitating the modular approach advocated above.
Notably, we generalise existing tall data splitting approaches \citep{Scott:2013wza, Neiswanger:2013ug} for independent, identically distributed data to other types of data.

Finally, we also develop an algorithm for fitting the Markov melded model in stages,
for both joining and splitting models.
This algorithm extends naturally that employed in \cite{Lunn:2013uo} and is closely related to those proposed in \cite{Liang:2007fi} and \cite{Tom:2010uf}.

The paper is organised as follows: in Section \ref{s:motivating-examples} we introduce some examples motivating this work; Section \ref{sec:integrating-two-sub} provides the conceptual framework underlying our approach; inferential and computational aspects of the approach are presented in Section \ref{sec:inference}; Section \ref{sec:examples} gives details and results for the motivating examples; we conclude with a discussion and suggestions for further work in Section \ref{sec:discussion}.

\section{Motivating examples}
\label{s:motivating-examples}

We motivate and demonstrate our framework for joining and splitting models with two examples, for which we provide here a brief high-level outline.
We show how Markov melding applies in each case in Section~\ref{s:motivating-examples-melding}; and  provide full details and results in Section~\ref{sec:examples}.
For both examples, as in the rest of the paper, we use directed acyclic graphs (DAGs) to represent the dependence structure between variables in a model (Figures~\ref{fig:flu-model-high-level} and \ref{fig:ecology-high-level}).
Each variable in the model is represented by a node with rectangular nodes denoting observed variables and links between the nodes indicating direct dependencies. Stochastic (distributional) dependencies are represented by solid lines and deterministic (logical) relationships by dashed lines.
The joint distribution of  all nodes is the product of the conditional distributions of each node given its direct parents, and conditional independence relationships can be read from the graph \citep{Lauritzen96}.

\subsection{Joining: A/H1N1 influenza evidence synthesis}
\label{s:motivating-example-synthesis-evidence}

\begin{figure}[t]
\centering
\subfloat[][ICU submodel]{
\begin{tikzpicture}[minimum width=0.75cm, minimum height = 0.75cm]
\node[circle, draw] (link) at (0, 0.5) {\(\cpar\)};
\node[draw] (fludata) at (-1.5, 0) {\(\fluicudata\)};
\node[circle, draw] (fluicugenericpara) at (-1.5, 2.25) {\(\fluicugenericpara\)};
\node[ellipse, draw] (fluicugenericposdata) at (0, 2.25) {\(\fluicugenericposdata\)};

\node (dummy1) at (-3.5, 0) {};
\node (dummy2) at (2, 0) {};

\draw[dashed,->] (fluicugenericpara) -- (link);
\draw[dashed,->] (fluicugenericposdata) -- (link);
\draw[->] (fluicugenericpara) -- (fludata);
\end{tikzpicture}
}
\hspace{2em}
\subfloat[][Severity submodel]{
\centering
\begin{tikzpicture}[minimum width=0.75cm, minimum height = 0.75cm]
\node[circle, draw] (link) at (0, 0) {\(\cpar\)};

\node[double, circle, draw] (fluna) at (0, 2.25) {\(\fluna\)};
\node[ellipse, draw] (fludetprob) at (1.5, 2.25) {\(\fludetprob\)};

\draw[->] (fluna) -- (link);
\draw[->] (fludetprob) -- (link);

\node (dummy1) at (-0.5, 0) {};
\node (dummy2) at (2, 0) {};
\end{tikzpicture}
}
\caption{High-level DAG representations of the influenza submodels.
The double circle denotes the (highly) informative prior for \(\fluna\), reflecting data from the full severity submodel that is omitted here.
Detailed DAGs of these submodels are shown in Figure~\ref{fig:flumodel}.}
\label{fig:flu-model-high-level}
\end{figure}
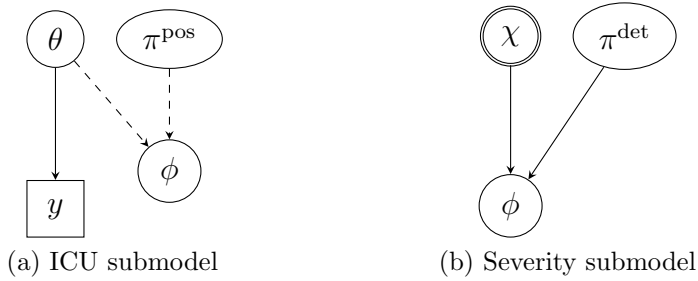

Public health responses to influenza outbreaks rely on knowledge of severity: the probability that an infection results in a severe event such as hospitalisation or death. One method to estimate severity is by combining estimates of cumulative numbers of severe events with estimates of cumulative numbers of infections obtained from synthesizing different data sources. This approach is adopted in \citet{Presanis:2014em} for the A/H1N1 pandemic, where information from intensive care units (ICU) is integrated with several other sources. Figure~\ref{fig:flu-model-high-level} provides a schematic representation of the submodels used for each evidence component. A crucial ingredient is the cumulative number of ICU admissions for  the A/H1N1 strain, $\chi$. A lower bound $\phi$ for $\chi$  is estimable through an immigration-death model governed by transition rates $\theta$, from time-dependent (weekly) prevalence data $y$ on suspected `flu cases in ICU (Figure~\ref{fig:flu-model-high-level}(a)). The $\phi$ of Figure~\ref{fig:flu-model-high-level}(a) is a deterministic function (a sum) of latent quantities involving $\theta$ and other parameters \(\fluicugenericposdata\). Indirect aggregate evidence on $\chi$ is also available from a severity submodel (Figure~\ref{fig:flu-model-high-level}(b)), whose complexity is summarised here by an informative prior on $\chi$. The lower bound $\phi$ is related to $\chi$ through a binomial model with probability parameter \(\fludetprob\).

The two submodels imply two different prior models for the link quantity \(\phi\).
A further complication is that the deterministic function connecting \(\phi\) to the ICU submodel parameters is a sum of products, which is not invertible, preventing the ICU submodel from being expressed conditional on \(\phi\).
\cite{Presanis:2014em} therefore transferred information between the two separate submodels via an approximate approach (see Section~\ref{sec:approx-methods} for details).
We show in this paper how Markov melding can be used to join the two submodels formally into a single joint model, making all the assumptions involved explicit.
We also explain the relationship between our approach and the approximate approach.

\subsection{Splitting: large ecology model}
\label{s:motivating-example-partitioning}
\begin{figure}[t]
\centering

\subfloat[][Joint model]{
\begin{tikzpicture}[minimum width=0.75cm, minimum height = 0.75cm]
\node[circle, draw] (link) at (0, 1.75) {\(\cpar\)};

\node[circle, draw] (productivityrate) at (2, 1.75) {\(\productivityrate\)};
\node[draw] (indexdata) at (2, 0) {\(\indexdata\)};

\node[draw] (recoverydata) at (-2, 0) {\(\recoverydata\)};
\node[circle, draw] (recoveryrate) at (-2, 1.75) {\(\recoveryrate\)};

\draw[->] (link) -- (indexdata);
\draw[->] (productivityrate) -- (indexdata);

\draw[->] (link) -- (recoverydata);
\draw[->] (recoveryrate) -- (recoverydata);
\end{tikzpicture}
}%
\hspace{3em}%
\subfloat[][The joint model split into two submodels]{
\begin{tikzpicture}[minimum width=0.75cm, minimum height = 0.75cm]
\node[circle, draw] (link1) at (1, 1.75) {\(\cpar\)};
\node[circle, draw] (link2) at (-1, 1.75) {\(\cpar\)};

\node[circle, draw] (productivityrate) at (3, 1.75) {\(\productivityrate\)};
\node[draw] (indexdata) at (3, 0) {\(\indexdata\)};

\node[draw] (recoverydata) at (-3, 0) {\(\recoverydata\)};
\node[circle, draw] (recoveryrate) at (-3, 1.75) {\(\recoveryrate\)};

\draw[->] (link1) -- (indexdata);
\draw[->] (productivityrate) -- (indexdata);

\draw[->] (link2) -- (recoverydata);
\draw[->] (recoveryrate) -- (recoverydata);

\draw [dashed] (0,-0.5) -- (0,2.5);

\node (dummy1) at (-4, 0) {};
\node (dummy2) at (4, 0) {};
\end{tikzpicture}
}
\caption{High-level DAG representations of the ecology models.
Detailed DAG representations of these models are shown in Figure~\ref{fig:ecology-model}.}
\label{fig:ecology-high-level}
\end{figure}
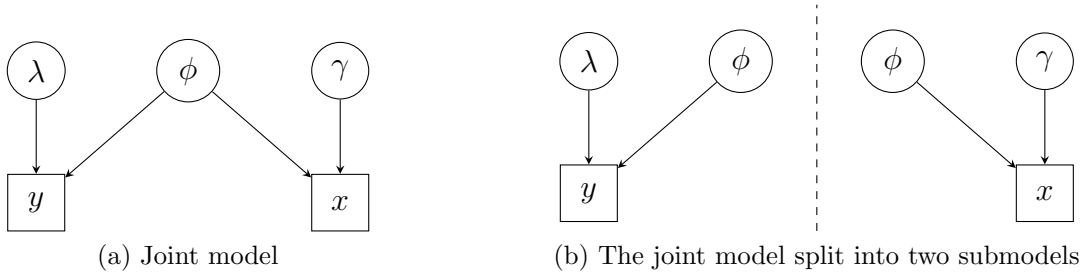

As an example of splitting a large DAG model we consider a joint model \citep{Besbeas:2002jw} for two distinct sources of data about British Lapwings ({\em Vanellus vanellus\/}).
These data sources are primarily collected to inform different aspects of studies of the birds: census-type data provide a measure of breeding population size, while mark-recapture-recovery data provide estimates of the annual survival probability of the birds via observations of the survival of uniquely marked individuals.
These data are related and a joint model allows inference to account for all information available.
In the joint Bayesian model of \cite{Brooks:2004wr} (Figure~\ref{fig:ecology-high-level}(a)), the mark-recapture-recovery data \(\recoverydata\) are modelled in terms of the recovery rate \(\recoveryrate\), and the survival rate \(\cpar\) for birds; and the census data \(\indexdata\) are modelled in terms of the survival rate \(\cpar\), and the productivity rate \(\productivityrate\) of adult female birds.
The joint model links the data sources using the common survival rate parameter~\(\cpar\).

\cite{Brooks:2004wr} considered fitting the census and mark-recapture-recovery models both separately and jointly using standard Markov chain Monte Carlo (MCMC) algorithms, but considering the joint model simultaneously is cumbersome and MCMC convergence is slow.
We describe in this paper how, through Markov melding, inference from such a joint model can be carried out in stages after splitting the model into two separate submodels (Figure~\ref{fig:ecology-high-level}(b)), circumventing the need to directly fit the joint model in a single MCMC procedure.
The multi-stage fitting process is more computationally efficient, and gives insight into the contribution of each submodel to the joint model.

\section{Conceptual framework}
\label{sec:integrating-two-sub}

\subsection{Joining models}
\label{sec:joining-models}

To combine probabilistic models in a principled way, we propose \emph{Markov melding} as an extension of a \emph{Markov combination}, which has been introduced by \citet{Dawid:1993fc} and discussed extensively with generalisations and applications in \citet{Massa:2010uf} and \citet{Massa:2017bz}.

Let $p$ denote either a probability distribution for discrete random variables or a probability density for continuous variables (we assume such a density exists).
In both cases we talk of $p$ interchangeably as a probability or probability distribution and we express conditional probabilities as \( p(\psi \mid \phi) = p(\psi,\phi)/p(\phi) \), where $p(\phi) > 0$. We will assume that when conditioning on a variable its distribution has support in the relevant region. For random variables \(X_{1}\), \(X_{2}\) and \(X_{3}\), \(X_{1} \ci X_{2} \mid X_{3}\) means that \(X_{1}\) and \(X_{2}\) are conditionally independent given \(X_{3}\).

\subsubsection{Markov combination}
\label{sec:markov-combination}

\citet{Dawid:1993fc} define the submodels $p_m(\phi,\psi_m,Y_m)$, $m=1,\ldots,M$, as {\em consistent\/} in the link parameter \(\cpar\) if the prior marginal distributions \(p_m(\cpar)= p(\phi) \) are the same for all $m$.
They define the {\em Markov combination} \(\pcomb\) of \(M\) consistent submodels as the joint model
\begin{equation}\begin{split}
\label{e:markov-combination}
\pcomb(\phi,\psi_1,\ldots,\psi_M,Y_1,\ldots,Y_M)
&=
p(\phi) \prod_{m=1}^M p_m(\psi_m,Y_m\mid \phi)\\
&=
\frac{\prod_{m=1}^M p_m(\phi,\psi_m,Y_m)}{p(\phi)^{M-1}}
\end{split}\end{equation}%
By construction, model~\eqref{e:markov-combination} assumes that the submodels are conditionally-independent: \((\psi_{m}, Y_{m}) \ci (\psi_{\ell}, Y_{\ell}) \given \cpar\) for $m\neq \ell$ (see Figure~\ref{fig:markov-combination}).
All prior marginal distributions and submodel-specific conditional distributions, given the link parameter, are preserved: \(\pcomb(\cpar, \psi_{\modelIndex}, \data_{\modelIndex}) = p_{\modelIndex}(\cpar, \psi_{\modelIndex}, \data_{\modelIndex})\) and  \(\pcomb(\psi_m, Y_m \mid \phi) = p_m(\psi_m, Y_m \mid \phi)\) for all \(\modelIndex\).
Furthermore, the model has maximal entropy among the set of distributions with this marginal preservation property, and so can be viewed as the least constrained among such distributions \citep{Massa:2010uf}.
However, only prior marginals are preserved in Markov combinations: the posterior distributions of \(\cpar\) and any \(\psi_{\ell}\) under the Markov combination model account for all data $Y_m$, $m=1,\ldots,M$, rather than just the submodel-specific data $Y_\ell$, and are not preserved.

\subsubsection{Markov melding}
\label{sec:markov-melding}

If the submodels are not consistent in their link parameter \(\cpar\), that is, if the prior marginal distributions $p_1(\phi)$, $\dots$, $\p_M(\cpar)$ of the link parameter differ, a Markov combination cannot be formed directly. However, the original submodels \(p_m(\phi,\psi_m,Y_m)\), \(\modelIndex = \modelFirst, \dots, \modelLast\), can be altered so that the marginals $p_1(\phi), \dots, \p_M(\cpar)$ for the link parameter become consistent. This is achieved by a procedure we term {\em marginal replacement\/}, where a new model \(\preplm(\phi,\psi_m,Y_m)\) is formed by replacing the marginal distribution $p_m(\phi)$ of \(\cpar\) in the original model \(p_m(\phi,\psi_m,Y_m)\) by a new marginal distribution 
\(\pagg(\cpar)\):
\begin{equation}\begin{split}
\label{e:marginal-replacement}
\preplm(\phi,\psi_m,Y_m)
&=
p_m(\psi_m,Y_m \mid \phi)\, \pagg(\phi)\\
&=
\frac{p_m(\phi,\psi_m,Y_m)}{p_m(\phi)}\pagg(\phi)
\end{split}\end{equation}
where the {\em pooled density\/} $\pagg(\phi)=\poolfun(p_1(\phi),\ldots,p_M(\phi))$ is a function \(\poolfun\) of the individual prior marginal densities. Here, and in what follows, we assume that such a pooled density exists, that $g$ has been chosen such that $\int \pagg(\phi)\,d\phi = 1$, and that \(\pagg\) reflects an appropriate summary of the individual marginal distributions (we discuss options below). 

Since $\preplm(\phi,\psi_m,Y_m)$, \(\modelIndex = \modelFirst, \dots, \modelLast\), are consistent in the link parameter $\phi$ (that is, they all have the same prior marginal $\pagg(\phi)$), we can form their Markov combination
\begin{equation}\begin{split}
\label{e:markov-melding}
\pmeld(\phi,\psi_1,\ldots,\psi_M,Y_1,\ldots,Y_M)
 &= \pagg(\phi) \prod_{m=1}^M \preplm(\psi_m,Y_m \mid \phi)\\
 &=\pagg(\phi) \prod_{m=1}^M \frac{p_m(\phi,\psi_m,Y_m)}{p_m(\phi)}
\end{split}\end{equation}
We term this construction {\em Markov melding\/} of the submodels $p_m(\phi,\psi_m,Y_m)$ with pooled density $\pagg(\phi) = g(p_1(\phi),\ldots,p_M(\phi))$, which amounts to applying the Markov combination~(\ref{e:markov-combination}) to submodels satisfying the consistency condition \emph{after} marginal replacement as in~(\ref{e:marginal-replacement})\footnote{Markov melding can also be seen as a form of super Markov combination in the sense of proposition 4.14 in \citet{Massa:2010uf} where either family $\cal F$ or $\cal G$ have been extended to comprise a density with the marginal on the link variable replaced by the pooled density.}.
The submodel-specific conditional distributions, given the link parameter, are preserved in the Markov melded model: \(\pmeld(\psi_m, Y_m \mid \phi) = p_m(\psi_m, Y_m \mid \phi)\) for all \(\modelIndex\).
However, in contrast to Markov combination, the prior marginal distributions \(p_m(\cpar, \psi_{\modelIndex}, \data_{\modelIndex})\) will, in general, not be preserved in the Markov melded model.
Once the new model $\pmeld(\phi,\psi_1,\ldots,\psi_M,Y_1,\ldots,Y_M)$ has been formed by Markov melding, posterior inference conditioning on the data $Y_1 = \obsdata_{1},\ldots,Y_M = \obsdata_{M}$ can be performed (see Section~\ref{sec:inference}).

By extending a similar argument used in \citet{Poole:2000ee}, it can be shown that marginal replacement has the attractive property that $\preplm$ minimises the Kullback-Leibler divergence $D_\mathrm{KL}$ of a distribution $q(\phi,\psi,Y)$ to $p_m(\phi,\psi_m,Y_m)$ under the constraint that the marginals on $\phi$ agree, $q(\phi) = \pagg(\phi)$:
\[ \preplm(\phi,\psi_m,Y_m) = \mbox{argmin}_q
  \{ D_\mathrm{KL}(q \parallel p_m) \mid
            q(\phi) = \pagg(\phi) \mbox{ for all $\phi$} \} \]
Marginal replacement can also be interpreted as a generalisation of Bayesian updating in the light of new information.
Details are provided in Supplementary Material~\ref{s:appendix-kl}.

\subsubsection{Markov melding with deterministic variables}
Care is required when some of the dependencies in a submodel are deterministic, as in the example in Section~\ref{s:motivating-example-synthesis-evidence}.
The considerations are identical to those for Bayesian melding \citep{Poole:2000ee}, where priors on the input and output of deterministic functions are combined.
Specifically, assume the $k$-dimensional link parameter $\phi$ is deterministically related to a $\ell$-dimensional parameter $\theta$, $k\leq \ell$, in a model $p(\phi,\theta,\psi,Y)$. The probability model is effectively given by $p(\theta,\psi,Y)$ and $\phi$ follows an induced distribution.
We assume $\phi$ is exclusively a deterministic function $\phi(\theta)$ of the parameter $\theta$.

To apply Markov melding, we need to ensure that the prior marginal distribution on \(\phi\) is well defined, and that we can apply marginal replacement to \(\phi = \phi(\theta)\).
We must assume that $\phi(\theta)$ is an invertible function or, in the case of $k < \ell$, that $\phi(\theta)$ can be expanded into an invertible function $\phi_e(\theta) = (\phi(\theta),t(\theta))$, with a $\ell-k$ dimensional deterministic function $t(\theta)$.
We denote the inverse function by $\theta(\phi,t)$.
The function $\phi_e$ induces a probability distribution on $(\phi,t,\psi,Y)$ which can be represented as
\begin{equation*}%
 p(\phi,t,\psi,Y) = p(\theta(\phi,t),\psi,Y)\, J_\theta(\phi,t)
\end{equation*}
 where $J_\theta(\phi,t)$ is the Jacobian determinant for the transformation $\theta(\phi,t)$.
The marginal distribution on $\phi$ can now be obtained as \( p(\phi) = \int p(\phi,t,\psi,Y) \, dt\, d\psi\, dY \).
We show in Supplementary Material~\ref{s:appendix-deterministic} that \(p(\phi)\) is independent of the chosen parametric extension~$t(\theta)$ and so is well defined, and that we can apply marginal replacement, as defined by~\eqref{e:marginal-replacement}, to replace \(p(\phi)\) with \(\pnew(\phi)\):
\begin{equation}\begin{split}
\label{e:marginal-replacement-deterministic}
\prepl(\theta,\psi,Y)
= \frac{p(\theta,\psi,Y)}
   {p(\phi(\theta))} \pagg(\phi(\theta))
\end{split}\end{equation}
Markov melding with \(\prepl(\theta,\psi,Y)\) can now be applied as in~\eqref{e:markov-melding}.

\subsubsection{Pooling marginal distributions}
The pooling function \(\poolfun\) determines the prior marginal distributions \(\pmeld(\cpar, \psi_{\modelIndex}, \data_{\modelIndex})\), which, in general, will not match those in the original submodels.
It must, therefore, be chosen subjectively, ensuring that the pooled density \(\pagg(\phi)\) appropriately represents prior knowledge of the link parameter \(\cpar\).
Various standard pooling functions have been suggested in the multiple expert elicitation literature \cite[see, for example,][]{Clemen:1999vy, OHagan:2006}.
The difference here is that we propose to pool prior marginal distributions of submodels, rather than directly-specified priors.
A simple option is {\em linear pooling\/}, %
\begin{equation*}%
\pagg(\phi)
=
\frac{1}{\normalising_{\mathrm{lin}}(\aggpar)}
\sum_{m = 1}^{M} \aggpar_m p_m(\phi),
\quad \normalising_{\mathrm{lin}}(\aggpar) = \int \sum_{m = 1}^{M} \aggpar_m p_m(\phi) \,d\phi
\end{equation*}
where $\aggpar=(\aggpar_1, \dots,\aggpar_M)^{\top}$, with $w_m \geq 0$ to weight the submodel priors.
An alternative is {\em log pooling\/},
\begin{equation*}%
\pagg(\cpar)
=
\frac{1}{\normalising_{\mathrm{log}}(\aggpar)}\prod_{m =1}^{M} p_m(\phi)^{\aggpar_m},
\quad \normalising_{\mathrm{log}}(\aggpar) = \int \prod_{m = 1}^{M} p_m(\phi)^{\aggpar_m} \,d\phi
\end{equation*}
with $\aggpar_m \geq 0$,
a logarithmic version of the linear pooling (for reasons why logarithmic pooling might be attractive see Supplementary Material~\ref{s:externally-bayesian-pooling}). A special case of log pooling is {\em product of experts (\poe{}) pooling\/} \citep{Hinton:2002ic} when $\aggpar_m=1$ for all~$m$
\begin{equation*}%
\pagg(\cpar)
=
\frac{1}{\normalising_{\mathrm{poe}}}\prod_{m = 1}^{M} p_m(\phi),
 \quad \normalising_{\mathrm{poe}} = \int \prod_{m=1}^{M} p_m(\phi)\,d\phi
\end{equation*}
in which equal weight is given to each submodel prior.
A further special case of linear or log pooling is {\em dictatorial pooling\/} \(\pagg(\cpar) = p_{m_0}(\phi)\) when one submodel $m_0$ is considered authoritative\footnote{Dictatorial pooling corresponds to left (or right) composition in the terminology of \cite{Massa:2010uf}, and for $M=2$ their upper Markov combination is the family comprising two distributions namely the two possible directions for dictatorial pooling.}.
We shall assume throughout this paper that the weights \(\aggpar\) are a fixed quantity, chosen subjectively, in contrast to some of the power prior literature \citep{Neuenschwander:2009}, where attempts have been made to treat the weight \(\aggpar\) as an unknown parameter.

\begin{figure}[t]
\centering
\includegraphics[width=0.9\linewidth]{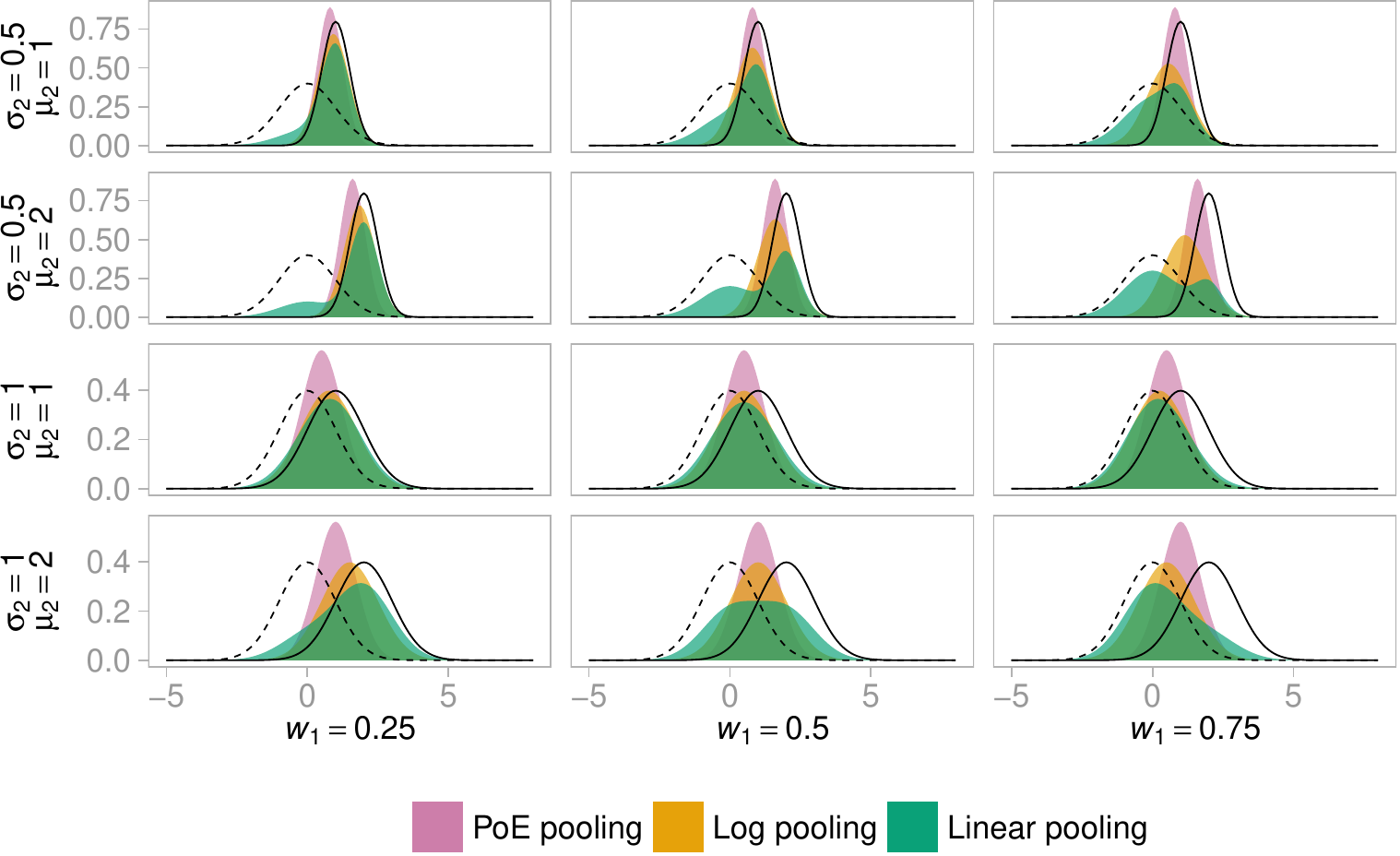}
\caption[Pooled densities under various pooling functions]{
Pooled densities under \poe{}, log and linear pooling, with \(\aggpar_{1} = 0.25\), 0.5 and 0.75 (and \(\aggpar_{2} = 1 - \aggpar_{1}\)), formed by pooling a \(\text{N}(0, 1)\) density (\tikz[baseline=-0.5ex]{ \draw [-, dashed,line width=0.75pt] (0,0) -- (3ex,0); }) and a  \(\text{N}(\mu_{2}, \sigma_{2}^{2})\) density (\tikz[baseline=-0.5ex]{ \draw [-, line width=0.75pt] (0,0) -- (3ex,0); }) with \(\mu_2 = 1, 2, 4\), and \(\sigma_2 = 0.5, 1\).
}
\label{fig:pooling}
\end{figure}

Figure~\ref{fig:pooling} shows the pooled density when combining two normal distributions under three different pooling functions with three choices of weights.
The \poe{} approach is arguably the least intuitive pooling function due to the rather concentrated combined distribution implied.
However, the required computation is greatly simplified (see Section~\ref{sec:inference}), and so if (and only if) \poe{} adequately represents prior beliefs then \poe{} pooling is an attractive option.
The choice of pooling function is particularly important when there is some disagreement between the priors, but if there is substantial conflict between submodel priors we do not recommend the use of Markov melding, as mentioned above.

\subsection{Splitting models}
\label{sec:splitting-with-markov-melding}

We may want to split up a larger model, for example, for computational efficiency or to understand the influence of each submodel on the joint model. In this case we want to split a large joint model \(\p(\phi,\psi_1,\ldots,\psi_M,Y_1,\ldots,Y_M)\) into \(\modelLast\) submodels $p_m(\phi,\psi_m,Y_m)$,
$m=1,\ldots,M$, in such a way that joining the submodels using Markov melding recovers the original, joint model.
If \((\psi_{m}, Y_{m}) \ci (\psi_{\ell}, Y_{\ell}) \given \cpar\) for $m\neq \ell$ in the original model, then suitable submodels are
\[
p_m(\phi,\psi_m,Y_m) = p(\psi_m,Y_m \mid \phi)p_m(\phi),
\qquad
\modelIndex = \modelFirst, \dots, \modelLast
\]
where \(p_1(\phi), \dots, p_{\modelLast}(\cpar)\) are new prior marginal distributions.
These marginal distributions and the pooling function \(\poolfun\) can chosen freely to enable efficient computation, provided that the pooled distribution $\pagg(\phi) = \poolfun(p_1(\phi), \allowbreak\dots, \allowbreak p_{\modelLast}(\phi))$ is the same as the original marginal distribution $p(\phi)$.
An obvious choice is \(p_m(\cpar) = p(\cpar)^{1/\modelLast}\) with \poe{} pooling, but there are many other options.
For example, \poe{} pooling is suitable for {\em any\/} factorisation of \(p(\cpar)\) into \(\modelLast\) factors.

Note that a splitting strategy based on Markov melding is suitable {\em only} if \((\psi_{m}, Y_{m}) \ci (\psi_{\ell}, Y_{\ell}) \given \cpar\) for $m\neq \ell$, that is, conditioning on the link variable $\phi$ makes the parts that are intended for splitting conditionally independent.

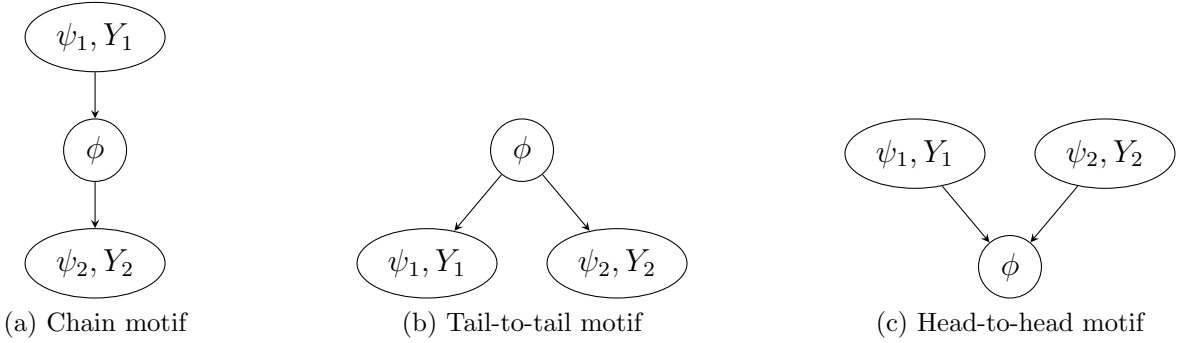
\begin{figure}[t]
\centering
\subfloat[][Chain motif]{
\begin{tikzpicture}[minimum width=0.75cm, minimum height = 0.75cm]
\node[circle, draw] (link) at (0, 1.5) {\(\cpar\)};
\node[ellipse, draw] (l1) at (0, 3) {\(\latent_{\modelFirst}, \data_{\modelFirst}\)};
\node[ellipse, draw] (l2) at (0, 0) {\(\latent_{\modelSecond}, \data_{\modelSecond}\)};

\node (dummy1) at (-1,0) {};
\node (dummy2) at (1,0) {};

\draw[->] (l1) -- (link);
\draw[->] (link) -- (l2);
\end{tikzpicture}
}
\hspace{4em}
\subfloat[][Tail-to-tail motif]{
\centering
\begin{tikzpicture}[minimum width=0.75cm, minimum height = 0.75cm]
\node[circle, draw] (link) at (0, 1.5) {\(\cpar\)};
\node[ellipse, draw] (l1) at (-1.25, 0) {\(\latent_{\modelFirst}, \data_{\modelFirst}\)};
\node[ellipse, draw] (l2) at (1.25, 0) {\(\latent_{\modelSecond}, \data_{\modelSecond}\)};

\draw[->] (link) -- (l1);
\draw[->] (link) -- (l2);
\end{tikzpicture}
}
\hspace{4em}
\subfloat[][Head-to-head motif]{
\centering
\begin{tikzpicture}[minimum width=0.75cm, minimum height = 0.75cm]
\node[circle, draw] (link) at (0, 0) {\(\cpar\)};
\node[ellipse, draw] (l1) at (-1.25, 1.5) {\(\latent_{\modelFirst}, \data_{\modelFirst}\)};
\node[ellipse, draw] (l2) at (1.25, 1.5) {\(\latent_{\modelSecond}, \data_{\modelSecond}\)};

\draw[->] (l1) -- (link);
\draw[->] (l2) -- (link);
\end{tikzpicture}
}
\caption{DAG representations of stylised situations where model splitting might be desirable.
Splitting the joint model is possible in (a) and (b), but not in (c).}
\label{fig:splitting}
\end{figure}

Figure~\ref{fig:splitting} shows a few stylised situations with \(\modelLast = 2\) where splitting for computational purposes might be desirable.
The joint distributions for all models is $p(\phi,\allowbreak \psi_1,\allowbreak \psi_2,\allowbreak Y_1,\allowbreak Y_2)$.
The model in Figure~\ref{fig:splitting}(a) can be split into $p_1(\phi,\psi_1,Y_1) = p(\phi,Y_1\mid\psi_1)p(\psi_1)$ and $p_2(\phi,\psi_2,Y_2) = p(\psi_2,Y_2\mid \phi)p_2(\phi)$, with a new prior distribution $p_2(\phi)$, which could be different and computationally simpler than $p(\phi)=\int p(\phi,Y_1\mid\psi_1)p(\psi_1)\,d\psi_1\,dY_1$.
Markov melding, with dictatorial pooling $\pagg(\phi) = p_1(\phi)$, results in
\begin{equation*}\begin{split}
 \pmeld(\phi,\psi_1,\psi_2,Y_1,Y_2) &= p_1(\phi)\,\frac{p_1(\phi,\psi_1,Y_1)}{p_1(\phi)}\frac{p_2(\phi,\psi_2,Y_2)}{p_2(\phi)} \\
 &= p(\phi,Y_1\mid\psi_1)\,p(\psi_1) p_2(\psi_2,Y_2 \mid \phi)
 = p(\phi,\psi_1,\psi_2,Y_1,Y_2),
\end{split}\end{equation*}
leading to the original model, regardless of the choice of \(p_2(\phi)\).
The case in Figure~\ref{fig:splitting}(b) is similar to the example in Section~\ref{s:motivating-example-partitioning} (see Section~\ref{sec:part-large-models} for a definition of splitting in this case).
Note that in Figures~\ref{fig:splitting}(a) and \ref{fig:splitting}(b), the dependencies between the nodes can include deterministic (logical) dependence, provided \((\latent_{1}, Y_{1}) \ci (\latent_{2}, Y_{2}) \given \cpar\), as usual.
The case in Figure~\ref{fig:splitting}(c) cannot be split into $p_1(\phi,\psi_1,Y_1)$ and $p_2(\phi,\psi_2,Y_2)$ by Markov melding model splitting because \((\latent_{1}, Y_{1}) \notci (\latent_{2}, Y_{2}) \given \cpar\).

\subsection{Markov melding in the motivating examples}
\label{s:motivating-examples-melding}

\subsubsection{Joining: A/H1N1 influenza evidence synthesis}
\label{sec:framework-synthesising-evidence}
Markov melding involves joining the ICU submodel (Figure~\ref{fig:flu-model-high-level}(a)) with density \(\pone(\cpar, \fluicugenericpara, \fluicugenericposdata, \allowbreak Y)\), where \(\cpar\) is a deterministic function of \(\fluicugenericpara\) and \(\fluicugenericposdata\), and the severity submodel (Figure~\ref{fig:flu-model-high-level}(b)) with density \(\ptwo(\cpar, \fluna, \fludetprob)\).
Replacing the marginal distribution of \(\cpar\) with pooled density \(\pagg(\cpar)\) in the ICU submodel, using~\eqref{e:marginal-replacement-deterministic}, and in the severity submodel, using~\eqref{e:marginal-replacement}, and then applying Markov melding, as in~\eqref{e:markov-melding}, results in
\begin{align}
\label{eq:influenza-melded}
\pmeld(\cpar, \fluicugenericpara, \fluicugenericposdata, \fluna, \fludetprob, Y)
&=
\pagg(\cpar)\,
\frac{\pone(\fluicugenericpara, \fluicugenericposdata, Y)}{\pone(\cpar(\fluicugenericpara, \fluicugenericposdata))}\,
\frac{\ptwo(\cpar, \fluna, \fludetprob)}{\ptwo(\cpar)}\,
\end{align}
where \(Y_1 = Y\), \(\psi_1 = \{\fluicugenericpara, \fluicugenericposdata\}\), \(Y_2 = \varnothing\) and \(\psi_2 = \{\fluna, \fludetprob\}\) in the notation of~(\ref{e:markov-melding}).

\subsubsection{Splitting: large ecology model}
\label{sec:part-large-models}\sloppy
The original, joint model (Figure~\ref{fig:ecology-high-level}(a)), with density $p(\phi,\recoveryrate,\productivityrate,Y,X)$, can be split into separate submodels (Figure~\ref{fig:ecology-high-level}(b)) with densities $\pone(\phi,\recoveryrate,Y)$ and $\ptwo(\phi,\productivityrate,X)$.
Provided the priors $p_1(\phi)$ and $p_2(\phi)$ in the separate submodels are such that  $\pagg(\phi)=\poolfun(p_1(\phi),p_2(\phi))$ equals the original marginal distribution $p(\phi)$, for some choice of pooling function \(\poolfun\), then Markov melding the submodels recovers the joint model:
\begin{equation*}\begin{split}
\pmeld(\cpar, \recoveryrate, \productivityrate,Y,X)
   &= \pagg(\phi)\,p_1(\recoveryrate,Y\mid \phi)\,p_2(\productivityrate, X \mid \phi)\\
   &= p(Y \mid \recoveryrate,\phi)\,p(\recoveryrate)\,p(X\mid \productivityrate,\phi)\,p(\productivityrate)\,p(\phi),
\end{split}\end{equation*}
with $\psi_1=\recoveryrate$, $Y_1=Y$, $\psi_2=\productivityrate$, $Y_2=X$ in the notation of~(\ref{e:markov-melding}).

\section{Inference and computation}
\label{sec:inference}\fussy

The joint posterior distribution, given data $Y_m=y_m$, \(\modelIndex = \modelFirst, \dots \modelLast\), under the Markov melded model in~\eqref{e:markov-melding} is
\begin{equation}
\pmeld(\phi,\psiall \mid \yall) \propto \pagg(\phi)\prod_{m=1}^M
    \frac{p_m(\phi,\psi_m, y_m)}{p_m(\phi)}
\label{eqn:full-joint-model}
\end{equation}
The degree of difficulty of inference for this posterior distribution depends on the specification of the submodels.
Our focus is settings in which, considered separately, each of the original collection of submodels is amenable to inference by standard Monte Carlo methods \citep[for example,][]{robertCasellaMCMC}.

In Section~\ref{sec:mh-samplers} we first consider a standard Metropolis-within-Gibbs sampler, but when the constituent submodels are complex, this sampler may be cumbersome and slow.
We thus propose a multi-stage Metropolis-within-Gibbs sampler, in which inference for the full Markov melded model is generated iteratively in stages, starting with standard inference on one of the submodels.
This latter sampling scheme enables a convenient modular approach to inference.
Both approaches, in general, require the marginal prior densities \(\p_{\modelIndex}(\cpar)\) of the link parameter under each submodel, \(\modelIndex = 1, \dots, \modelLast\), which will not usually be analytically tractable.
In Section~\ref{sec:estimating-link-marginals} we discuss approaches to estimating these densities, although there is no need to estimate them if \poe{} pooling is chosen.
In Section~\ref{sec:approx-methods} we show how approximate approaches, such as those used by \cite{Presanis:2014em}, relate to the Markov melded model.

\subsection{Metropolis-Hastings samplers}
\label{sec:mh-samplers}

A general Metropolis-Hastings sampler for the posterior distribution \eqref{eqn:full-joint-model} can be constructed in the usual way.
Candidate values $(\phi^\star,\psiallstar)$ for each parameter of the Markov melded model are drawn from a proposal distribution $q(\phi^\prop,\psiallstar \mid \phi,\allowbreak \psiall)$, based on the current values $(\phi,\psiall)$ of the Markov chain.
The candidate values are accepted with probability $\min(1,\mhprob)$, where $\mhprob$ is in the form
 \[ \mhprob = \frac{R(\phi^\prop,\psiallstar,\phi,\psiall)}
{R(\phi,\psiall,\phi^\prop,\psiallstar)}
\]
where the {\em target-to-proposal density ratio\/} is
\begin{equation}
\label{e:ttpdr}
\begin{split}%
R(\phi^\prop,&\psiallstar,\phi,\psiall) \\
&=\pagg(\phi^\prop)
\prod_{m=1}^M\frac{p_m(\phi^\prop,\psi_m^\prop,y_m)}{p_m(\phi^\prop)}
\times
\frac{1}{q(\phi^\prop,\psiallstar \mid \phi,\psiall)}
\end{split}\end{equation}

\subsubsection{Metropolis-within-Gibbs sampler}
A particular form of the above general sampler is a Metropolis-within-Gibbs sampling scheme \citep{Muller:1991wq}, in which samples are drawn from the full conditional distribution of each latent parameter \(\latent_{\modelFirst}, \dots, \latent_{\modelLast}\), and then the link parameter \(\cpar\) in turn.

\paragraph{Latent parameter updates}
Markov melding does not introduce any extra complexities in sampling the parameters \(\latent_{\modelIndex}\) in each submodel \(\modelIndex = \modelFirst, \dots, \modelLast\) (conditional on the link parameter \(\cpar\)) beyond those inherent to the original submodels, considered separately.
Typically, they can be sampled using standard algorithms.
For instance, a Metropolis-Hastings algorithm, in which we draw a candidate value \(\latent_{\modelIndex}^{\prop}\) from a proposal distribution \(\pprop(\latent_{\modelIndex}^{\prop} \given \latent_{\modelIndex})\) based upon the current value \(\latent_{\modelIndex}\), will be feasible whenever the corresponding algorithm is feasible for estimation of the posterior distribution of the \(\modelIndex\)\textsuperscript{th} submodel alone.
In this case, since terms involving marginal densities for the link parameter $\cpar$ in~\eqref{eqn:full-joint-model} cancel, the target-to-proposal density ratio in~\eqref{e:ttpdr} simplifies to
\begin{equation*}
R(\phi,\psionestar,\phi,\psiall)
=
p_m(\cpar,\latent_m^\prop,y_{\modelIndex})
\times
\frac{
1
}{
\pprop(\latent_{\modelIndex}^{\prop} \given \latent_{\modelIndex})
}
\end{equation*}
This target-to-proposal density ratio is identical to that required for a Metropolis-Hastings update for the parameter \(\latent_{\modelIndex}\), conditional on the link parameter \(\cpar\), when the \(\modelIndex\)\textsuperscript{th} submodel alone is the target distribution.

\paragraph{Link parameter updates}
To update the link parameters, a candidate value \(\cpar^{\prop}\) is drawn from an appropriate proposal distribution \(\pprop(\cpar^{\prop} \given \cpar)\), based upon the current value \(\cpar\), and is accepted according to the target-to-proposal density ratio
\begin{equation}%
\label{e:mwg-link}
R(\phi^\prop,\psiall,\phi,\psiall)
=
\pagg(\phi^\prop)
\prod_{m=1}^M\frac{p_m(\phi^\prop,\psi_m,y_m)}{p_m(\phi^\prop)}
\times
\frac{
1
}{
\pprop(\cpar^{\prop} \given \cpar)
}
\end{equation}
When the prior marginal distributions \(p_m(\cpar)\) or \(\pagg(\cpar)\) are not analytically tractable, we propose to use an approximation \(\widehat{p}_{\modelIndex}(\cpar)\) in their place, calculated using the methods described in Section~\ref{sec:estimating-link-marginals}.
Note that, under \poe{} pooling, the terms involving the marginal distributions for the link parameter \(\cpar\) cancel in~\eqref{e:mwg-link}, leaving
\begin{equation*}%
R(\phi^\prop,\psiall,\phi,\psiall)
=
\prod_{m=1}^Mp_m(\phi^\prop,\psi_m,y_m)
\times
\frac{
1
}{
\pprop(\cpar^{\prop} \given \cpar)
}
\end{equation*}
removing the need to estimate the marginal prior distribution for the link parameter \(\cpar\).

\subsubsection{Multi-stage Metropolis-within-Gibbs sampler}
\label{sec:multi-stage-mwg}

When the constituent submodels are complex, an alternative, multi-stage approach may be computationally preferable to the Metropolis-within-Gibbs sampler.
The multi-stage approach generalises the two stage approach in \cite{Lunn:2013uo}.
We assume a factorisation of the pooled prior \( \pagg(\phi) = \prod_{m=1}^M p_{\mathrm{pool},m}(\phi) \).
A default factorisation for any pooling function sets $p_{\mathrm{pool},m}(\phi)=\pagg(\phi)^{1/M}$, but there may be more computationally-efficient factorisations.
For example, when \poe{} pooling is used, the factorisation with $p_{\mathrm{pool},m}(\phi) = p_m(\phi)$ is more computationally efficient, as we describe below.
The aim then is to sample, iteratively in stages $\stage = 1,\ldots,M$, from
\begin{equation}\begin{split}\label{e:iterative-melded}
\p_{\mathrm{meld},\stage}(\phi,\psi_1,\ldots,\psi_\stage\mid y_1,\ldots,y_\stage)
\propto \prod_{m=1}^{\stage}\left(\frac{p_m(\phi,\psi_m , y_m)}{p_m(\phi)}
p_{\mathrm{pool},m}(\phi)\right)
\end{split}\end{equation}
Since \(\p_{\mathrm{meld},\modelLast}(\phi,\psi_1,\ldots,\psi_\modelLast\mid y_1,\ldots,y_\modelLast) = \pmeld(\phi,\psi_1,\ldots,\psi_\modelLast\mid y_1,\ldots,y_\modelLast)\), after \(M\) stages the samples obtained reflect the posterior distribution~\eqref{eqn:full-joint-model} of the full Markov melded model.
Note that each $p_m(\phi)$ (and thus also $\pagg(\phi)$) can be estimated from the submodels in advance and independently of the following sampling scheme, as we describe in Section~\ref{sec:estimating-link-marginals}.

\paragraph{Stage 1.}

We obtain \(\mcLast_1\) samples \((\cpar^{(\mcIndex,1)}, \psi_1^{(\mcIndex,1)})\), \(\mcIndex = \mcFirst, \dots, \mcLast_1\), drawn from $\p_\mathrm{meld,1}(\phi, \allowbreak\psi_1 \mid y_1)$.
The most appropriate method for obtaining such samples depends on the nature of the submodel \(\pone(\cpar, \latent_{\modelFirst}, \data_{\modelFirst})\); typically, standard Monte Carlo methods, such as MCMC, will be suitable.

\paragraph{Stage $\stage$.}

After we have sampled up to stage~$\stage-1$ from~(\ref{e:iterative-melded}), we construct a Metropolis-within-Gibbs sampler for stage $\stage$ for the parameters \((\cpar, \latent_{1},\ldots, \latent_{\stage})\) given data $(y_1,\ldots,y_{\stage})$.
The parameter~\(\latent_{\stage}\) is updated, conditional on the link parameter~\(\cpar\) and parameters $\psiallpart$ using a standard algorithm, such as a Metropolis-Hastings sampler, with a target-to-proposal density ratio
\[ R(\phi,\psi_1,\ldots,\psi_{\stage}^\star,\phi,\psi_1,\ldots,\psi_{\stage})
 = p_{\stage}(\phi,\psi_{\stage}^\star,y_{\stage})\times
      \frac{1}{q(\psi_{\stage}^\star\mid \phi,\psi_{\stage})} \]
We use the samples from stage~$\stage-1$ as a proposal distribution when updating the parameters \(\psi_1,\ldots,\psi_{\stage}\) and the link parameter~\(\cpar\).
Specifically, we draw an index~\(\mcIndexRandom\) uniformly at random from \(\{\mcFirst, \dots, \mcLast_{\stage-1}\}\), and set \((\phi^\star,\psiallstarpart) = (\phi^{(d,\stage-1)},\psi_1^{(d,\stage-1)},\ldots,\psi_{\stage-1}^{(d,\stage-1)})\), so that
\begin{equation*}\begin{split}
(\phi^{(d,\stage-1)},\psi_1^{(d,\stage-1)},\ldots,\psi_{\stage-1}^{(d,\stage-1)})
  \sim \p_{\mathrm{meld},\stage-1}(\phi,\psiallpart\mid \yallpart)
\end{split}\end{equation*}
The attraction of this particular proposal distribution is the resulting cancellation of likelihood terms for the first $\stage-1$ submodels in the target-to-proposal density ratio,
\begin{equation}\begin{split}\label{e:two-stage-R}
R(\phi^\prop,\psi_1^\star,\ldots,\psi_{\stage-1}^\star,\psi_{\stage},
     \phi,\psi_1,\ldots,\psi_{\stage-1},\psi_{\stage})
=\frac{p_{\stage}(\phi^\prop,\psi_{\stage},y_{\stage})}{p_{\stage}(\phi^\star)}
p_{\mathrm{pool},\stage}(\phi^\prop)
\end{split}\end{equation}
meaning this update step can be performed quickly. Once sampling in stage \(\stage\) has converged, samples \((\cpar^{(\mcIndex,\stage)}, \psi_1^{(\mcIndex,\stage)},\ldots,\psi_{\stage}^{(\mcIndex,\stage)})\), \(\mcIndex = \mcFirst, \dots, \mcLast_{\stage}\),
are obtained for use in stage \(\stage+1\).

The density ratio~(\ref{e:two-stage-R}) does not depend on parameters \(\psi_1,\ldots,\psi_{\stage-1}\), so if interest focuses entirely on the parameters \((\cpar, \psi_{\stage})\) then \(\psi_1,\ldots,\psi_{\stage-1}\) can be ignored in stage \(\stage\) of the multi-stage sampling: they do not need to be monitored or updated by the sampling algorithm.
The multi-stage sampler is nevertheless still sampling from the joint target distribution $\p_{\mathrm{meld},\stage}(\phi,\psi_1,\ldots,\psi_{\stage}\mid y_1,\ldots,y_{\stage})$.
Stage~$\stage$ is influencing the acceptance or rejection of samples of $\p_{\mathrm{meld},\stage-1}(\phi,\psiallpart\mid \yallpart)$ from the previous stage, thus adjusting this distribution according to the requirements of the joint model.

In general, evaluation of ratio~(\ref{e:two-stage-R}) requires estimates of the prior marginal distribution of the link parameter under the $\stage$\textsuperscript{th} submodel, which can be obtained as described in Section~\ref{sec:estimating-link-marginals}.
However, if \poe{} pooling is used and $p_{\mathrm{pool},m}(\phi) = p_m(\phi)$, \(m = 1, \dots, \modelLast\), the ratio simplifies to \(R(\phi^\prop,\psi_1^\star,\ldots,\psi_{\stage-1}^\star,\psi_{\stage}, \phi,\psi_1,\ldots,\psi_{\stage-1},\psi_{\stage}) = p_{\stage}(\phi^\prop,\psi_{\stage},y_{\stage})\), meaning that no estimates of the marginal distribution are required.

\subsection{Estimating marginal distributions}
\label{sec:estimating-link-marginals}
The prior marginal densities \(\p_{\modelIndex}(\cpar)\) of the link parameter under each of the \(\modelLast\) submodels are central to Markov melding, and in particular are required to evaluate the acceptance probability of proposals within the MCMC samplers we proposed above.
However, these marginals are not generally analytically tractable, except when the prior distribution \(\p_{\modelIndex}(\cpar)\) is directly-specified as a standard, tractable distribution, such as when \(\cpar\) appears as a founder node in a DAG representation of the submodel.
When not available analytically, we can estimate the marginal density \(\p_{\modelIndex}(\cpar)\) for each submodel \(\modelIndex\) by kernel density estimation \citep{henderson2015applied} with samples drawn from \(\p_{\modelIndex}(\cpar)
=
\iint
\p_{\modelIndex}(\cpar, \latent_{\modelIndex}, \data_{\modelIndex}) \,
d \latent_{\modelIndex} \,
d \data_{\modelIndex}\) by standard (forward) Monte Carlo.
Care is required if \(\cpar\) has high dimension because the curse of dimensionality applies to kernel density estimation; see Section~\ref{sec:discussion} for further discussion.

\subsection{Normal two-stage approximation method}
\label{sec:approx-methods}
Approximate approaches for joining submodels are widely used in applied research.
In this section, we show that approximate inference for the Markov melded model formed by joining  submodels (with \poe{} pooling) can be produced using a standard normal approximation approach.

Consider the case when the Markov melded model \(\pmeld(\phi,\psi_1,\psi_2,Y_1,Y_2)\) is formed by joining \(\modelLast = 2\) submodels \(\pone(\phi, \psi_1, Y_1)\) and \(\ptwo(\phi, \psi_2, Y_2)\).
Suppose that $\psi_1$ is not a parameter of interest in the posterior distribution, so that it can be integrated over
 \begin{equation}\begin{split}\label{e:joint-prob-integrated}
 \pmeld(\phi,\psi_2, Y_1,Y_2)
     &= \int \pmeld(\phi,\psi_1,\psi_2,Y_1,Y_2) \,d\psi_1\\
     &= \pagg(\phi) \int p_1(\psi_1,Y_1 \mid \phi)\,d\psi_1\,\,p_2(\psi_2,Y_2 \mid \phi)\\
     & = \pagg(\phi)\, p_1(Y_1\mid \phi)\,p_2(\psi_2,Y_2 \mid \phi)
\end{split}\end{equation}
An approximate two stage sampler that mimics the multi-stage sampler (above) can then be constructed for this marginal distribution.

\begin{description}
\item{\textbf{Stage 1}} Fit the submodel $p_1(\phi,\psi_1,Y_1)$ to obtain posterior samples from $p_1(\phi\mid Y_1)$, and approximate the posterior by a (multivariate) normal distribution with mean $\widehat{\mu}$ and covariance $\widehat{\Sigma}$.
With \(p_N\) denoting the probability density function for a (multivariate) normal distribution,
 \[ p_1(\phi \mid Y_1) \approx p_N(\phi \mid \widehat{\mu},\widehat{\Sigma})
 = p_N(\widehat{\mu} \mid \phi ,\widehat{\Sigma}) .\]

\item{\textbf{Stage 2}} Since $p_1(\phi,Y_1) \propto p_1(\phi \mid Y_1) \approx p_N(\widehat{\mu} \mid \phi ,\widehat{\Sigma})$,
we obtain an approximation for~(\ref{e:joint-prob-integrated}) by replacing \(p_1(\phi \mid Y_1)\) by \(p_N(\widehat{\mu} \mid \phi ,\widehat{\Sigma})\).
 \begin{equation*}\begin{split}%
 \pmeld(\phi,\psi_2, Y_1,Y_2)
    & \propto \pagg(\phi)\, \frac{p_1(\phi \mid Y_1)}{p_1(\phi)}\frac{p_2(\phi,\psi_2,Y_2)}{p_2(\phi)}\\
    & \approx \pagg(\phi)\, \frac{p_N(\widehat{\mu} \mid \phi ,\widehat{\Sigma})}{p_1(\phi)}\frac{p_2(\phi,\psi_2,Y_2)}{p_2(\phi)}\\
\end{split}\end{equation*}
\end{description}
This two-stage approximate approach is commonly used in practice (see Section~\ref{sec:discussion}) in the form \(\pmeld(\phi,\psi_2, Y_1,Y_2) \approx c\,p_N(\widehat{\mu} \mid \phi ,\widehat{\Sigma})\, p_2(\phi,\psi_2,Y_2)\), with $c$ a data dependent constant.
In this case the likelihood of the second submodel is modified by a factor $p_N(\widehat{\mu}\mid \phi,\widehat{\Sigma})$ (in a DAG representation a dependency of the constant $\widehat{\mu}$ on $\phi$ and the constant $\widehat{\Sigma}$ is added).
This approach can be viewed as approximate Markov melding with \poe{} pooling, in which one submodel is represented by a normal approximation.

If, instead of \poe{} pooling, one wishes to regard the marginal $p_2(\phi)$ on the link variable $\phi$ as authoritative and thus fully retain it, dictatorial pooling $\pagg(\phi)=p_2(\phi)$ leads to the variant
 \begin{equation*}\begin{split}%
 \pmeld(\phi,\psi_2, Y_1,Y_2)
    & \propto \frac{p_1(\phi \mid Y_1)}{p_1(\phi)}p_2(\phi,\psi_2,Y_2)
     \approx
\frac{p_N(\phi \mid \widehat{\mu} ,\widehat{\Sigma})}
{p_N(\phi \mid \widehat{\mu}_0 ,\widehat{\Sigma}_0)}
p_2(\phi,\psi_2,Y_2) \\
    &\propto
p_N(\phi \mid \mu_c ,\Sigma_c)\,
p_2(\phi,\psi_2,Y_2)
\end{split}\end{equation*}
where $\widehat{\mu}_0$ and $\widehat{\Sigma}_0$ are an estimate of the mean and covariance of the prior marginal $p_1(\phi)$, which can be obtained at stage one in parallel to the posterior by sampling from the prior submodel, and
\begin{equation*}%
    \Sigma_c^2 = \left(\widehat{\Sigma}^{-1} - {\widehat{\Sigma}_0}^{-1}\right)^{-1},\quad
    \mu_c = \Sigma_c
      \left(\widehat{\Sigma}^{-1}\widehat{\mu}
  - \widehat{\Sigma}_0^{-1}\widehat{\mu}_0\right).
\end{equation*}
Adjusting according to \(\widehat{\mu}_{0}\) and \(\widehat{\Sigma}_{0}\) removes the prior \(\pone(\cpar)\) from approximate joint model.

\section{Results}
\label{sec:examples}

\subsection{Joining: A/H1N1 influenza evidence synthesis}
\label{sec:h1n1-example}
\begin{figure}[t]
\centering
\subfloat[][ICU submodel]{
\begin{tikzpicture}[minimum width=0.75cm, minimum height = 0.75cm]
\node[circle, draw] (link) at (0, 1.5) {\(\cpar_{\fluage}\)};
\node[ellipse, draw] (flupos) at (1, 3) {\(\flupos_{\fluage, \flutimeset}\)};
\node[ellipse, draw] (fluposmin) at (1, 4.5) {\(\fluposmin_{\fluage, \flutimetwoset}\)};
\node[draw] (fluypos) at (3, 3) {\(\fluypos_{\fluage, \flutimetwoset}\)};
\node[draw] (flunpos) at (3, 4.5) {\(\flunpos_{\fluage, \flutimetwoset}\)};

\node[ellipse, draw] (flulambda) at (-1, 3) {\(\flulambda_{\fluage, \flutimeset}\)};
\node[circle, draw] (flugamma) at (-1, 4.5) {\(\flugamma_{\fluage}\)};

\node[ellipse, draw] (fluicu) at (-2, 1.5) {\(\flumean_{\fluage, \flutimeset}\)};
\node[draw] (fludata) at (-2, 0) {\(\fluicudata_{\fluage, \flustartofweekset}\)};

\node[circle, draw] (fluexitrate) at (-3, 3) {\(\fluexitrate_{\fluage}\)};
\node[circle, draw] (fluexitratealpha) at (-4, 6) {\(\fluexitratealpha\)};
\node[circle, draw] (fluexitratebeta) at (-2, 6) {\(\fluexitratebeta\)};

\draw[dashed, ->] (flupos) -- (link);
\draw[->] (fluposmin) -- (flupos);
\draw[->] (fluposmin) -- (fluypos);
\draw[->] (flunpos) -- (fluypos);

\draw[dashed, ->] (flulambda) -- (link);

\draw[->] (flugamma) -- (flulambda);

\draw[dashed, ->] (flulambda) -- (fluicu);

\draw[->] (fluicu) -- (fludata);

\draw[dashed, ->] (fluexitratealpha) -- (fluexitrate);
\draw[dashed, ->] (fluexitratebeta) -- (fluexitrate);
\draw[dashed, ->] (fluexitrate) -- (fluicu);

\node[draw, rectangle, rounded corners,
      fit={(link) (flupos) (fluposmin) (fluypos) (flunpos) (flulambda) (flugamma) (fluicu) (fludata) (fluexitrate)}] (plate) {};
\node[font = \footnotesize, node distance=0, inner sep=0pt, below left=-17.5pt and 5pt of plate.south east] {\(\fluage = 1, 2\)};
\end{tikzpicture}
}
\hspace{5em}
\subfloat[][Severity submodel]{
\centering
\begin{tikzpicture}[minimum width=0.75cm, minimum height = 0.75cm]
\node[circle, draw] (link) at (0, 0) {\(\cpar_{\fluage}\)};
\node[double, circle, draw] (fluna) at (0, 1.5) {\(\fluna_{\fluage}\)};
\node[ellipse, draw] (fludetprob) at (1.75, 1.5) {\(\fludetprob\)};

\node[draw, rectangle, rounded corners,
      fit={(link) (fluna) (-0.75,-1)}] (plate) {};
\node[font = \footnotesize, node distance=0, inner sep=0pt, below left=-17.5pt and 5pt of plate.south east] {\(\fluage = 1, 2\)};

\node (dummy1) at (-0.5,0) {};
\node (dummy2) at (2,0) {};

\draw[->] (fluna) -- (link);
\draw[->] (fludetprob) -- (link);
\end{tikzpicture}
}
\caption{
DAG representations of the submodels of A/H1N1 influenza.
Repeated variables are enclosed by a rounded rectangle, with the label denoting the range of repetition.
For simplicity the time domain is suppressed: parameters with subscripts \(\flutimeset\), \(\flustartofweekset\) and \(\flutimetwoset\) are collections of parameters across the time range denoted by the subscript.
For example, \(\fluicudata_{\fluage, \flustartofweekset} = \{\fluicudata_{\fluage, \flutime} : \flutime \in \flustartofweekset\}\).
}
\label{fig:flumodel}
\end{figure}
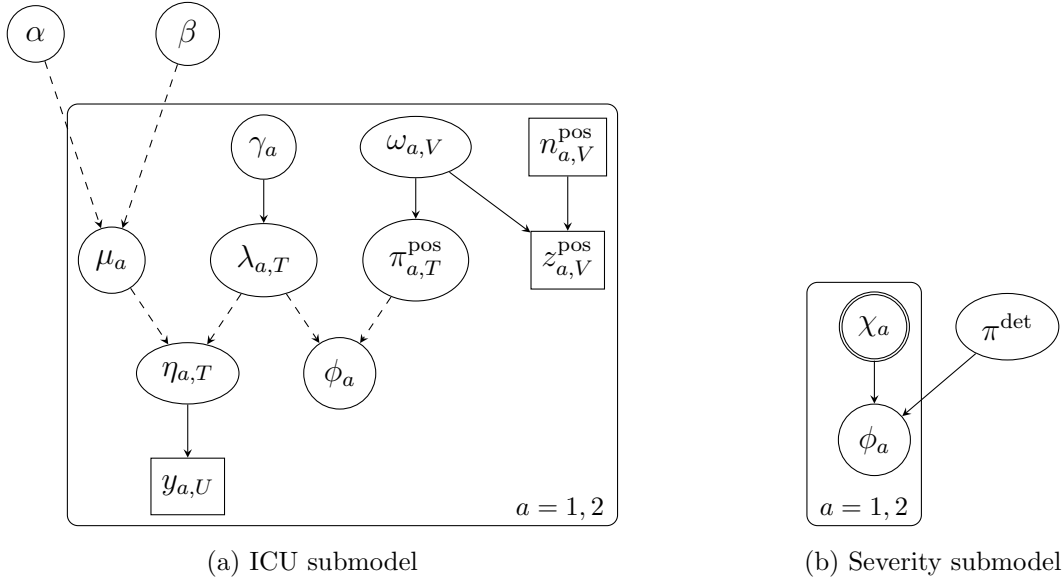

Figure~\ref{fig:flumodel} shows DAG representations of the two submodels outlined in Section~\ref{s:motivating-example-synthesis-evidence}.

\subsubsection{ICU submodel}
The main data source in the ICU submodel is prevalence-type data from the Department of Health's Winter Watch scheme \citep{DoHWW:2011}, which records the total number of patients in all ICUs in England with {\em suspected\/} pandemic A/H1N1 influenza infection.
Weekly observations \(\fluicudata_{\fluage, \flutime}\) taken at days \(\flutime \in \flustartofweek = \{8, 15, 22, \dots, 78\}\) for age group \(\fluage \in \{1, 2\}\) (children and adults respectively) are available between December 2010 and February 2011. To estimate the link parameter \(\cpar = (\cpar_{\fluage}) = (\cpar_{1}, \cpar_{2})\), that is, the cumulative number of ICU admissions over the period of observation \(\flutime \in \flutimeset = \{1, \dots, 78\}\), from such prevalence data requires an immigration-death model for the system of ICU admission and exits from ICU.
Assume that new ICU admissions follow an inhomogeneous Poisson process with rate \(\flulambda_{\fluage, \flutime}\) at time \(\flutime\), and the length of stay in ICU is exponentially distributed with rate \(\fluexitrate_{\fluage}\).
Then the number of patients admitted up to time \(\flutime\) who are still present in ICU at time \(\flutime\) follows a thinned inhomogeneous Poisson process and the observed number of prevalent patients is \(\fluicudata_{\fluage, \flutime} \distributedas \dpoi(\flumean_{\fluage, \flutime})\), \(\fluage \in \{1, 2\}\), \(\flutime \in \flustartofweek\), with expectation, under a discretised formulation with daily time steps, given by \(\flumean_{\fluage, \flutime} = \sum_{\flutimeint = 1}^{\flutime} \flulambda_{\fluage, \flutimeint} \exp\{-\fluexitrate_{\fluage}(\flutime - \flutimeint)\}\), \(\flutime \in \flutimeset\).
We assume \(\flumean_{\fluage, 1} = 0\) to enforce the assumption that no patients with suspected `flu were in ICU a week before observations began.

The product of the expected new admissions of suspected cases \(\flulambda_{\fluage, \flutime}\) and the proportion positive for A/H1N1 \(\flupos_{\fluage, \flutime}\) gives the expected number of {\em confirmed\/} new admissions on day \(\flutime\).
The link parameter \(\cpar_{\fluage}\) is the uninvertible sum of these products over time:
\begin{equation*}
\cpar_{\fluage}
=
\sum_{\flutime \in \flutimeset}
\flupos_{\fluage, \flutime}
\flulambda_{\fluage, \flutime},
\qquad
\fluage = 1, 2.
\end{equation*}
We model the proportion positive \(\flupos_{\fluage, \flutime}\) using weekly virological positivity data from the sentinel laboratory surveillance system Data Mart \citep{PHEdataMart:2014}, which records the number \(\fluypos_{\fluage, \flutimetwo}\) of A/H1N1-positive swabs out of the total number \(\flunpos_{\fluage, \flutimetwo}\) tested during week \(\flutimetwo \in \flutimetwoset = \{1,\dots,11\}\) in age group \(\fluage \in \{1, 2\}\). 
We assume a uniform prior \(\flupos_{\fluage, \flutime} \distributedas \dunif(\fluposmin_{\fluage, \flutimetwo}, 1)\), \(\flutime \in \flutimeset\), for the true positivity, where \(\flutimetwo = 1\) for \(\flutime = 1, \dots, 14\) and \(\flutimetwo = \lfloor (\flutime - 1)/7 \rfloor\) for \(\flutime = 15, \dots, 78\), and where the lower bound \(\fluposmin_{\fluage, \flutimetwo}\) is informed by a binomial model for the positivity data: \(\fluypos_{\fluage, \flutimetwo} \distributedas \dbin(\flunpos_{\fluage, \flutimetwo}, \fluposmin_{\fluage, \flutimetwo})\), \(\flutimetwo \in \flutimetwoset\).
For the expected new admissions \(\flulambda_{\fluage, \flutime}\), 
we assume a random-walk prior with \(\log(\flulambda_{\fluage, 1}) \distributedas \dunif(0, 250)\) and \(\log(\flulambda_{\fluage, \flutime}) \distributedas \dnorm(\log(\flulambda_{\fluage, \flutime - 1}), \flugamma_{\fluage}^{-2})\) for \(\flutime = 2, \dots, 78\), with \(\flugamma_{\fluage} \distributedas \dunif(0.1, 2.7)\).
For the length of ICU stays we assume constant age-group specific exit rates \(\fluexitrate_{1} = \exp(-\fluexitratealpha)\) and \(\fluexitrate_{2} = \exp(-\{\fluexitratealpha + \fluexitratebeta\})\), with \(\fluexitratealpha \distributedas \dnorm(2.7058, 0.0788^2)\) and \(\fluexitratebeta \distributedas \dnorm(-0.4969, 0.2048^2)\) 
\citep{Presanis:2014em}.

\subsubsection{Severity submodel}
We consider a simplified version of the full, complex severity submodel in \cite{Presanis:2014em}.
The Winter Watch ICU data are only available for a portion of the time of the `third wave' of the A/H1N1 pandemic, and so the cumulative number of confirmed new admissions \(\cpar_{\fluage}\) from the ICU submodel is a lower bound for the true number \(\fluna_{\fluage}\) of ICU admissions during the third wave.
We thus assume \(\cpar_{\fluage} \distributedas \dbin(\fluna_{\fluage}, \fludetprob)\), \(\fluage \in \{1, 2\}\), where \(\fludetprob\) is the age-constant detection probability, to which we assign a \(\dbeta(6, 4)\) prior.
We incorporate the remaining evidence in the full severity submodel of \cite{Presanis:2014em} via  informative priors \(\fluna_{1} \distributedas \dlnorm(4.93, 0.17^2)\) and \(\fluna_{2} \distributedas \dlnorm(7.71, 0.23^2)\).

\subsubsection{Markov melded model}\fussy
We joined the submodels as in~\eqref{eq:influenza-melded}.
We considered linear and log pooling with pooling weight \(\aggpar_{1} = 0.25, 0.5 \text{ and } 0.75\) (and \(\aggpar_{2} = 1 - \aggpar_{1}\)), and \poe{} pooling.

\begin{figure}[t]
\centering
\includegraphics[width=\linewidth]{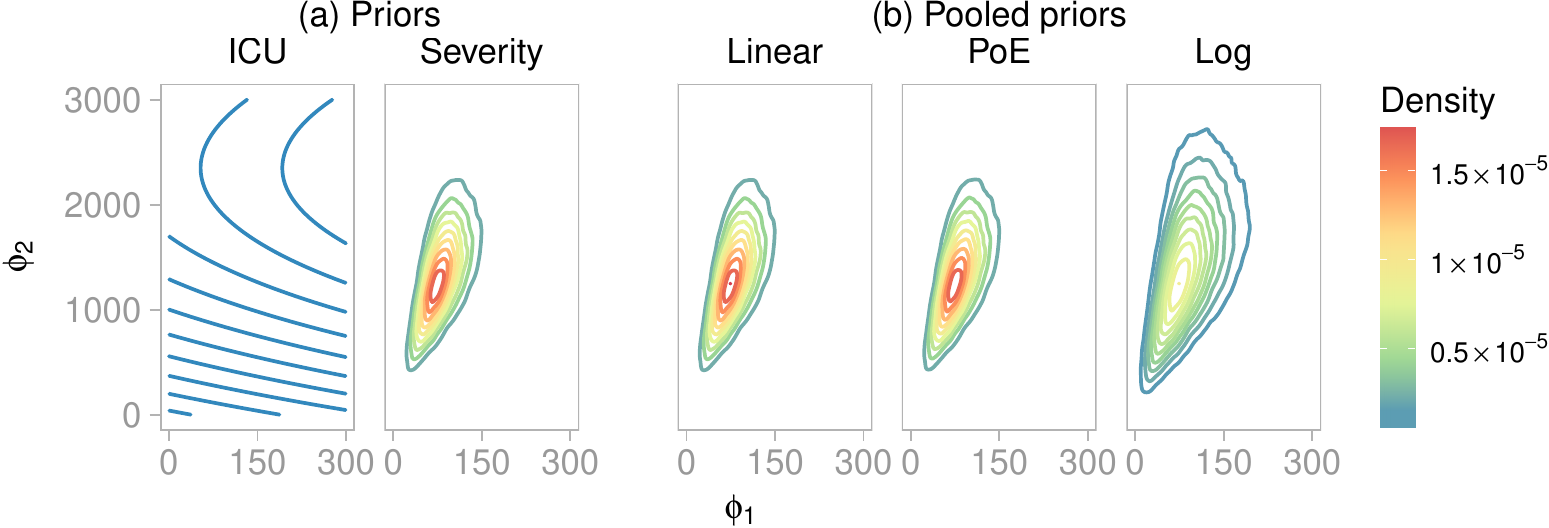}
\caption{\label{fig:h1n1-priors} Prior distributions for \(\cpar_{\fluage}\), the cumulative number of {\em confirmed\/} new admissions in age group \(\fluage\), in the A/H1N1 influenza evidence synthesis: (a) under the ICU and severity submodels; (b) pooled priors under three pooling functions with \(\aggpar_{1} = \aggpar_{2} = 0.5\).
}
\end{figure}

We estimated the marginal priors for \(\cpar = (\cpar_{1}, \cpar_{2})\) under the ICU and severity submodels using kernel density estimation with a bivariate \(t\)-distribution kernel, using \(5 \times 10^4\) independent draws, sampled from the corresponding submodel by forward Monte Carlo.
The marginal priors are shown in Figure~\ref{fig:h1n1-priors}(a).
Note that the ICU submodel prior for \(\cpar\) is extremely flat, whereas the severity submodel prior is concentrated on a small part of the parameter space.
The combined density using each of the pooling functions (with \(\aggpar_{1} = \aggpar_{2} = 0.5\)) is shown in Figure~\ref{fig:h1n1-priors}(b).
Linear and \poe{} pooling in this case lead to similar densities, whereas the log pooling prior is more dispersed.

\begin{figure}[t]
\centering
\includegraphics[width=\linewidth]{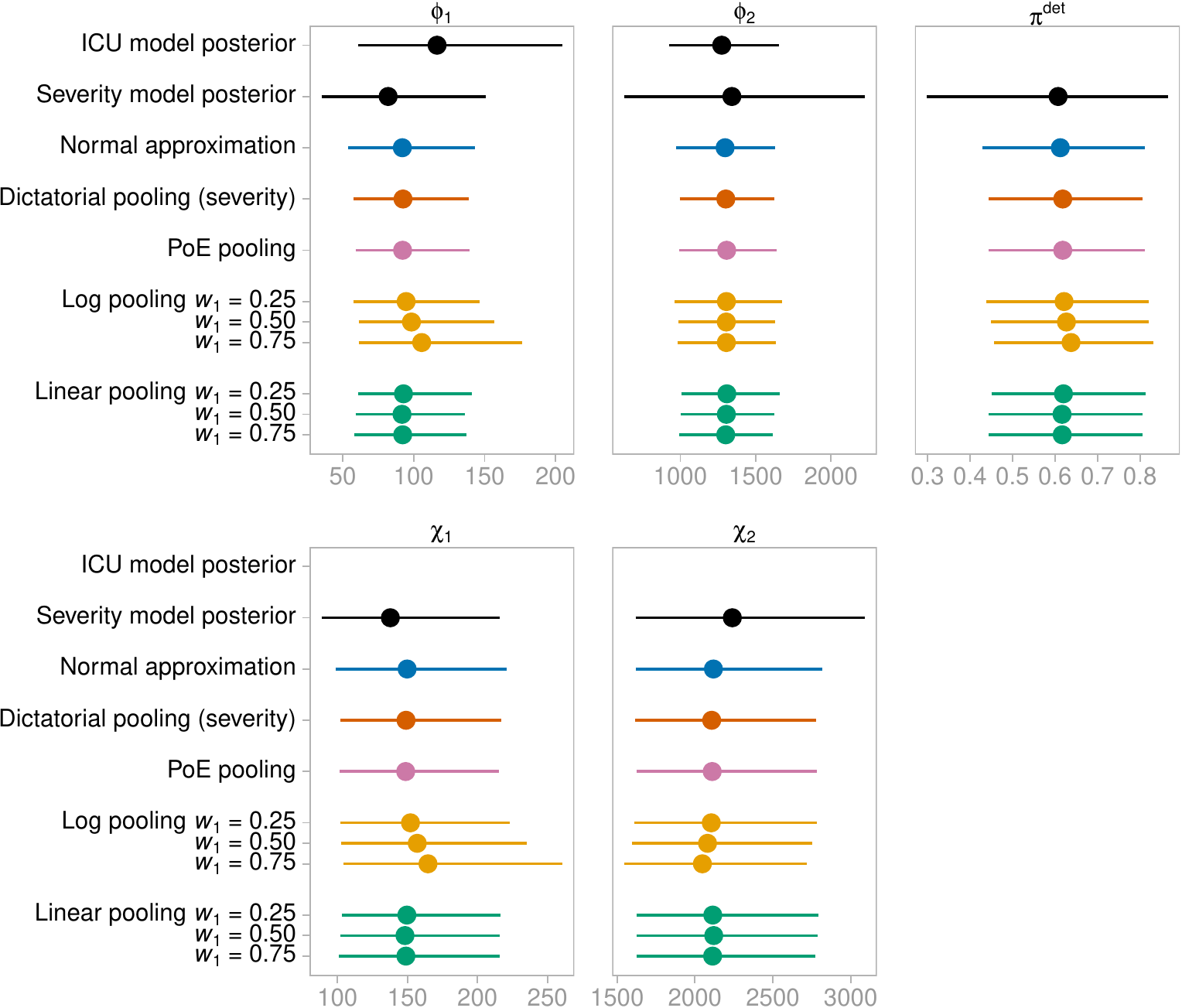}

\caption{\label{fig:h1n1-posteriors} Medians and 95\% credible intervals for: the posterior distribution for the link parameters \(\cpar_{1}\) and \(\cpar_{2}\) under the ICU submodel; the prior distribution for each parameter under the severity submodel; the posterior distribution for each parameter according to the normal approximation and the Markov melded model under each pooling function.
The \(x\)-axis shows number of individuals, except for \(\fludetprob\), which shows probabilities.}
\end{figure}

We then estimated, in stage one, the posterior distribution of the link parameter \(\cpar\) under the ICU submodel alone.
We drew 5 million iterations from the ICU submodel using JAGS \citep{plummer2015jags}, retaining every 100\textsuperscript{th} iteration, after discarding \(5 \times 10^4\) iterations as burn-in.
In stage two, for the Markov melded models under linear, log and \poe{} pooling, we drew \(2 \times 10^6\) samples using the multi-stage Metropolis-within-Gibbs sampler, with the first \(10^4\) samples discarded as burn-in.

Figure~\ref{fig:h1n1-posteriors} shows the results.
There is a notable reduction in uncertainty in the posteriors from Markov melding compared to the ICU submodel posterior, especially in \(\phi_{1}\), demonstrating the benefit of joining the submodels.
In the adult age group (\(\fluage = 2\)), the Markov melding results are robust to the choice of pooling function and pooling weight: the likelihood from the ICU submodel dominates over the pooled prior.
There is considerable agreement between the various approaches in the child age group (\(\fluage = 1\)) as well, although the choice of pooling weight has some influence on the upper tail under log pooling.
As anticipated by Section~\ref{sec:approx-methods}, the normal approximation (fitted using OpenBUGS) and PoE pooling posteriors are close, due to the near normality of the ICU posterior distribution.

\subsection{Splitting: large ecology model}
Figure~\ref{fig:ecology-model} is a DAG representation of the full, joint model outlined in Section~\ref{s:motivating-example-partitioning}.

\begin{figure}[t]
\centering

\begin{tikzpicture}[minimum width=0.75cm, minimum height = 0.75cm]

\draw[dotted] (-8.25, -2.5) rectangle (-4.5, 4.25);
\draw[dotted] (8.25, -2.5) rectangle (4.5, 4.25);

\node (recoverylabel) at (-7, -1.75) [text width=2cm]{Recovery submodel};
\node (indexlabel) at (7, -1.75) [text width=2cm,align=right]{Census submodel};

\node[circle, draw] (linkadult) at (1.575, 1.75) {\(\survivalrate_{\brooksadult}\)};
\node[circle, draw] (linkchild) at (-1.575, 1.75) {\(\survivalrate_{\brookschild}\)};

\node[ellipse, draw] (adultalpha) at (1.575, 3.5) {\(\brooksalphaadult\)};
\node[ellipse, draw] (adultbeta) at (3.425, 3.5) {\(\brooksbetaadult\)};
\node[ellipse, draw] (childalpha) at (-3.425, 3.5) {\(\brooksalphachild\)};
\node[ellipse, draw] (childbeta) at (-1.575, 3.5) {\(\brooksbetachild\)};
\node[draw] (freezing) at (0, 3.5) {\(\freezing\)};

\draw[dashed, ->] (adultalpha) -- (linkadult);
\draw[dashed, ->] (adultbeta) -- (linkadult);
\draw[dashed, ->] (childalpha) -- (linkchild);
\draw[dashed, ->] (childbeta) -- (linkchild);
\draw[dashed, ->] (freezing) -- (linkadult);
\draw[dashed, ->] (freezing) -- (linkchild);

\node[circle, draw] (prodratealpha) at (5.5, 3.5) {\(\brooksalpha_{\productivityrate}\)};
\node[circle, draw] (prodratebeta) at (7.5, 3.5) {\(\brooksbeta_{\productivityrate}\)};
\node[circle, draw] (productivityrate) at (6.5, 1.75) {\(\productivityrate\)};
\node[circle, draw] (truecounts) at (5.5, 0) {\(\truecounts\)};
\node[circle, draw] (indexsd) at (7, 0) {\(\indexsd^{2}\)};
\node[draw] (indexdata) at (5.5, -1.75) {\(\indexdata\)};

\node[draw] (recoverydata) at (-5.5, -1.75) {\(\recoverydata\)};
\node[circle, draw] (obsprob) at (-5.5, 0) {\(\obsprob\)};

\node[circle, draw] (recovratealpha) at (-7.5, 3.5) {\(\brooksalpha_{\recoveryrate}\)};
\node[circle, draw] (recovratebeta) at (-5.5, 3.5) {\(\brooksbeta_{\recoveryrate}\)};
\node[circle, draw] (recoveryrate) at (-6.5, 1.75) {\(\recoveryrate\)};

\draw[->] (linkadult) -- (truecounts);
\draw[->] (linkchild) -- (truecounts);

\draw[dashed, ->] (prodratealpha) -- (productivityrate);
\draw[dashed, ->] (prodratebeta) -- (productivityrate);
\draw[->] (productivityrate) -- (truecounts);
\draw[->] (indexsd) -- (indexdata);
\draw[->] (truecounts) -- (indexdata);

\draw[dashed, ->] (recovratealpha) -- (recoveryrate);
\draw[dashed, ->] (recovratebeta) -- (recoveryrate);
\draw[->] (recoveryrate) -- (obsprob);
\draw[->] (linkadult) -- (obsprob);
\draw[->] (linkchild) -- (obsprob);
\draw[->] (obsprob) -- (recoverydata);
\end{tikzpicture}

\caption{DAG representation of the joint ecology model.
The recovery and census submodels are connected via the common parameter \(\cpar = (\brooksalphachild, \brooksalphaadult, \brooksbetachild, \brooksbetaadult)\).
For simplicity the time domain is suppressed: \(\recoverydata\), \(\indexdata\), \(\obsprob\), \(\truecounts\), \(\recoveryrate\), \(\survivalrate_{\brookschild}\), \(\survivalrate_{\brooksadult}\), \(\productivityrate\) and \(\freezing\) represent the collection of {\em all\/} quantities sharing the same variable name.
For example, \(\truecounts = \{\truecounts_{\brooksagegroupindex, \brookst} : \brooksagegroupindex \in \{\brookschild, \brooksadult\},\, \brookst = 3, \dots, 36\}\).
}

\label{fig:ecology-model}
\end{figure}

\subsubsection{Mark-recapture-recovery data}
Mark-recapture-recovery data \(\recoverydata_{\brookst_{1}, \brookst_{2}}\) record the number of ringed birds released before May in year \(\brookst_{1} = 1, \dots, 35\), and recovered (dead) in the 12 months up to April in year \(\brookst_{2} = \brookst_{1} + 1, \dots, 36\).
The years correspond to observations for releases from 1963 (\(\brookst = 1\)) to 1997 and recoveries from 1964 to 1998.
The number of birds \(\recoverydata_{\brookst_{1}, 37}\) released in year \(\brookst_{1}\) and never recovered is also available.
We assume
\begin{equation*}
(\recoverydata_{\brookst_{1}, \brookst_{1} + 1},
\dots,
\recoverydata_{\brookst_{1}, 37})
\distributedas
\dmult(
\obsprob_{\brookst_{1}, \brookst_{1} + 1},
\dots,
\obsprob_{\brookst_{1}, 37}
),
\qquad
\brookst_{1} = 1, \dots, 35.
\end{equation*}
We model the probability \(\obsprob_{\brookst_{1}, \brookst_{2}}\) of recovery in year \(\brookst_{2}\) following release in year \(\brookst_{1}\) in terms of the recovery rate \(\recoveryrate_{\brookst}\), and the survival rates \(\survivalrate_{\brookschild,\brookst}\) and \(\survivalrate_{\brooksadult, \brookst}\) for immature (1 year old) and breeding (2 years or older) birds, respectively, up to April of year \(\brookst\):
\begin{equation*}
\obsprob_{\brookst_{1}, \brookst_{2}}
=
\begin{cases}
\recoveryrate_{\brookst_{2}}
(1 - \survivalrate_{\brookschild, \brookst_{2}})
&
\brookst_{1} = 1, \dots, 35, \, \brookst_{2} = \brookst_{1} + 1
\\
\recoveryrate_{\brookst_{2}}
\survivalrate_{\brookschild, \brookst_{1} + 1}
(1 - \survivalrate_{\brooksadult, \brookst_{2}})
&
\brookst_{1} = 1, \dots, 34, \, \brookst_{2} = \brookst_{1} + 2
\\
\recoveryrate_{\brookst_{2}}
\survivalrate_{\brookschild, \brookst_{1} + 1}
(1 - \survivalrate_{\brooksadult, \brookst_{2}})
\prod_{\brooksyearindex = \brookst_{1} + 2}^{\brookst_{2} - 1}
\survivalrate_{\brooksadult, \brooksyearindex}
&
\brookst_{1} = 1, \dots, 33, \, \brookst_{2} = \brookst_{1} + 3, \dots, 36
\end{cases}
\end{equation*}
The recovery rate is the probability that a bird that dies in year \(\brookst\) is recovered.
The probability of a bird released in year \(\brookst_{1}\) being never recovered is \(\obsprob_{\brookst_{1}, 37} = 1 - \sum_{\brooksyearindex = \brookst_{1} + 1}^{36} \obsprob_{\brookst_{1}, \brooksyearindex}\).

\subsubsection{Census data}
We assume that the observed census-type data \(\indexdata_{\brookst}\), which are available for 1965 (\(\brookst = 3\)) to 1998, account for only breeding birds and that there is no emigration.
We model the census data via the true number of breeding females  \(\truecounts_{\brooksadult, \brookst}\) and immature females \(\truecounts_{\brookschild, \brookst}\), and the productivity rate \(\productivityrate_{\brookst}\), the average number of female offspring per breeding female in year~\(\brookst\), which could be greater than 1.
Specifically we assume for \(\brookst = 3, \dots, 36\)
\begin{align*}
\indexdata_{\brookst}
&\distributedas
\dnorm(
\truecounts_{\brooksadult, \brookst},
\indexsd^{2}
)
\\
\truecounts_{\brookschild, \brookst}
&\distributedas
\dpoi(
\truecounts_{\brooksadult, \brookst - 1}\productivityrate_{\brookst - 1}\survivalrate_{\brookschild, \brookst}
)
\\
\truecounts_{\brooksadult, \brookst}
&\distributedas
\dbin(
\truecounts_{\brookschild, \brookst - 1}
+
\truecounts_{\brooksadult, \brookst - 1}
,
\survivalrate_{\brooksadult, \brookst}
),
\end{align*}
with the observation variance \(\indexsd^2\) assumed constant.

\subsubsection{Regression models and prior distributions}

We model the parameters \(\survivalrate_{\brooksagegroupindex, \brookst}\), \(\recoveryrate_{\brookst}\) and \(\productivityrate_{\brookst}\) with regression models, with \(\freezing_{\brookst}\) denoting the (observed) number of frost days in year \(\brookst\).
\begin{align*}
\logit(\survivalrate_{\brooksagegroupindex, \brookst})
&=
\brooksalphaagegroup
+
\brooksbetaagegroup
\freezing_{\brookst}
\qquad
\brooksagegroupindex = \brookschild, \brooksadult
\\
\logit(\recoveryrate_{\brookst})
&=
\brooksalpha_{\recoveryrate}
+
\brooksbeta_{\recoveryrate}
\brookst
\\
\log(\productivityrate_{\brookst})
&=
\brooksalpha_{\productivityrate}
+
\brooksbeta_{\productivityrate}
\brookst
\end{align*}
We place lognormal priors on the number of immature females \(\truecounts_{\brookschild,2}\) and breeding females \(\truecounts_{\brooksadult,2}\) in the year prior to our data series, with scale parameter \(1\) and location parameters \(\truecounts_{\brookschild} = 200\) and \(\truecounts_{\brooksadult} = 1000\) respectively.
We assume \(\indexsd^{2} \distributedas \dinvgamma(0.001, 0.001)\) a priori, and independent \(\dnorm(0, 10^2)\) prior distributions for all 8 regression parameters (\(\brooksalphachild, \brooksalphaadult, \brooksalpha_{\recoveryrate}, \brooksalpha_{\productivityrate}, \brooksbetachild, \brooksbetaadult, \brooksbeta_{\recoveryrate}, \brooksbeta_{\productivityrate}\)).

\subsubsection{Results}
We split the joint model, as described in Section~\ref{sec:part-large-models}, into two components: the mark-recapture-recovery submodel and the census submodel.
Denote by \(\brookssharedpara = (\survivalrate_{\brookschild}, \survivalrate_{\brooksadult},\allowbreak \brooksalphachild,\allowbreak \brooksalphaadult,\allowbreak \brooksbetachild, \brooksbetaadult)\) the parameters shared by both submodels and by \(\brooksrecoverypara = (\obsprob, \recoveryrate, \brooksalpha_{\recoveryrate}, \brooksbeta_{\recoveryrate})\) the parameters specific to the recovery submodel.
Under both the mark-recapture-recovery submodel (stage one) and the census submodel (stage two), we use independent normal priors, with mean 0 and standard deviation \(\sqrt{200}\), for each component \(\brooksalphachild\), \(\brooksalphaadult\), \(\brooksbetachild\) and \(\brooksbetaadult\) of the link parameter.
These priors were chosen so that \poe{} pooling of these priors results in the original prior for the link parameters under the joint model.

In stage one we drew samples from the posterior distribution \(\pone(\brooksrecoverypara, \brookssharedpara \given \recoverydata)\) under the recovery submodel, and retained these samples for use as a proposal distribution in stage two, in which we drew samples under the full joint model.
In stage one, we drew \(2.5 \times 10^5\) MCMC iterations from the posterior distribution of the mark-recapture-recovery submodel, taking 7 hours on a single core of an Intel Xeon E5-2620 2.0GHz CPU.
In stage two, we discarded all but every 100\textsuperscript{th} iteration, leaving \(2.5 \times 10^5\) MCMC iterations for inference.
This took \(6\frac{1}{2}\) hours.

\begin{figure}[t]
\centering
\includegraphics[width=\linewidth]{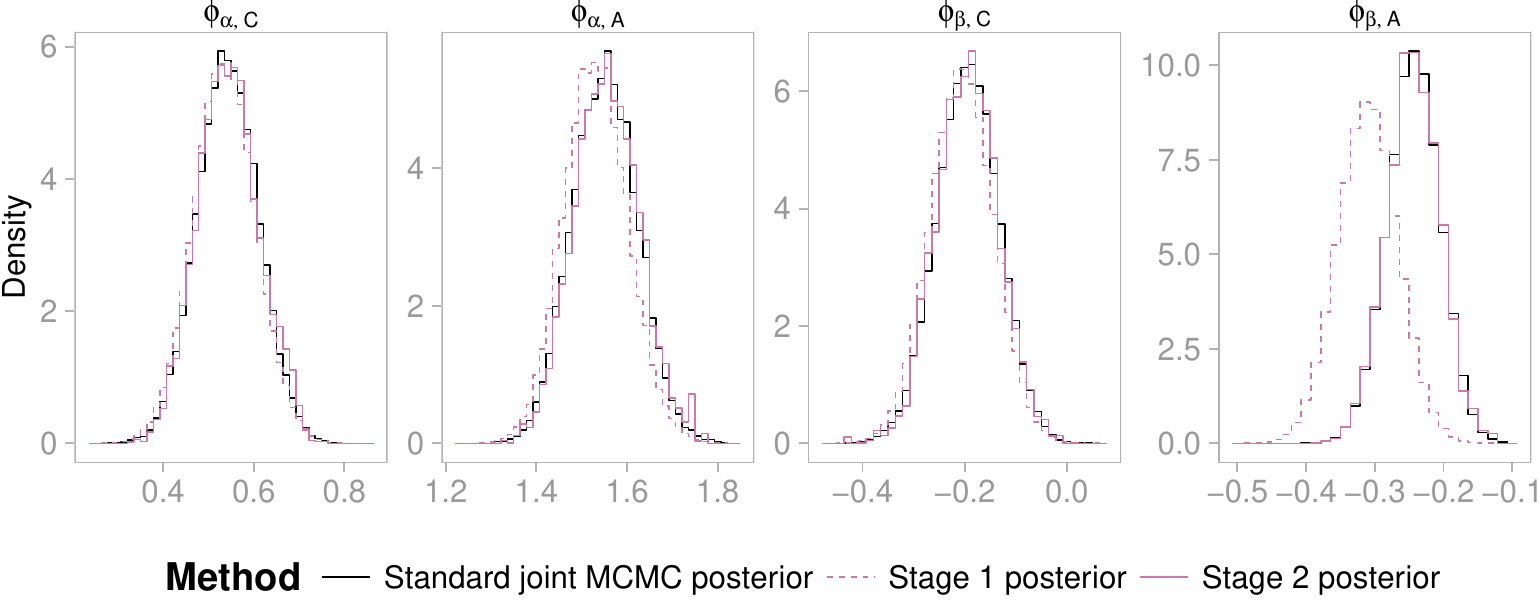}

\caption{\label{fig:ecology-results-densities} Histograms of the posterior densities of the link parameters \(\brooksalphachild\), \(\brooksalphaadult\), \(\brooksbetachild\) and \(\brooksbetaadult\) under the recovery submodel (Stage 1), and under the full joint model, as estimated by Stage 2 of the two stage sampler and by a standard MCMC sampler for the joint model.}
\end{figure}

Figure~\ref{fig:ecology-results-densities} shows the results.
We compare the two-stage estimates to the estimates of the joint distribution based upon \(6 \times 10^5\) MCMC iterations (retaining every 10\textsuperscript{th} iteration) drawn using a standard (one stage) MCMC sampler, which took 22 hours to run in OpenBUGS.
We regard these results as the `gold standard' that we aim to match with the two stage sampling approach.
The components of the link parameters \(\brooksalphachild\) and \(\brooksbetachild\) corresponding to the immature birds have posterior distributions that closely agree under the joint model and mark-recapture-recovery submodel alone, but there are differences in the parameters corresponding to mature birds.
In particular there is a sizeable difference for the regression parameter \(\brooksbetaadult\), which is estimated to be notably higher under the joint model than under the mark-recapture-recovery submodel alone.
The two-stage approach accurately captures this shift (Figure~\ref{fig:ecology-results-densities}, right-hand panel).
The similarity of \(\brooksalphachild\), \(\brooksalphaadult\) and \(\brooksbetachild\) in the stage 1 posterior (recovery model) and the stage 2 posterior implies that the census submodel contains little information about these parameters.
In contrast, the census submodel does contain information about the regression parameter \(\brooksbetaadult\) describing the relationship between the survival rate of adult birds and the number of frost days.
The census information suggests that \(\brooksbetaadult\) should be less negative than implied by the recovery information, implying that adult survival rate decreases only slightly in harsher winters.

\section{Further work and discussion}
\label{sec:discussion}

We have presented a unifying view and a generic method for joining and splitting probabilistic submodels that share a common variable.
We have extended the notion of Markov combination to the case where prior marginal distributions in each submodel need not be identical, enabling a principled approach to joining models in realistic applied settings, assuming that there is not strong conflict between evidence components and that it is reasonable to assume that the submodels are conditionally independent.
We also introduced a computational algorithm that allows inference for submodels to be efficiently conducted in stages, when considering either joining or splitting models.
The remainder of this section discusses related work, computational issues and alternative approaches.

\subsection{Related work}
The key idea for a melding approach can be attributed to \citet{Poole:2000ee}, but their presentation focuses on a limited set of models and is tied up with a deterministic link parameter~$\phi$.
This slightly obscures the key issues that we present more generally in Section~\ref{sec:integrating-two-sub}, where we clearly separate issues relating to marginal replacement from issues related to deterministic transformation of random variables.
A further influence is the work on decomposable graphical models \citep{Dawid:1993fc}, where a key concept is the separator, a subset of variables that splits the model into two parts that are independent conditional on the separator.
Separators correspond to link variables in Markov melding.
The rich literature on decomposable graphs and corresponding algorithms, such as junction tree algorithms \citep{Lauritzen96}, suggests extensions of Markov melding to a series of link variables (separators) for joining several submodels into chain or tree formations.

Evidence synthesis models \citep{EddyEtAl1992,Jackson:2009km,Albert:2011em,Commenges:2012gv} often employ the approximate approach of summarising the results of a first-stage submodel via a Gaussian or other distribution, for use in a second-stage submodel as a likelihood term.
We demonstrated in Section~\ref{sec:approx-methods} that this approach is an approximation to Markov melding under \poe{} pooling, therefore justifying the approximation.
Similar approximations are widely used in standard and network meta-analysis \cite[for example,][]{Hasselblad:1992iw,Ades:2006tw,Welton:2008iy}.
Similarly, in more general hierarchical models, splitting models to make inference faster or easier has previously been considered \citep{Liang:2007fi, Tom:2010uf, Lunn:2013uo}.
In this setting, posterior inference is first obtained from independent unit-specific submodels, with flat, independent priors replacing all hierarchical priors in the joint model.
Inference for the joint model is recovered in stage two through Markov melding of these unit-specific submodels with dictatorial pooling, so that only the hierarchical prior is reflected in the final results.
This can make cross-validation more convenient {\protect\citep{Goudie:2015kg}}.
Splitting models into conditionally independent components at a set of separator or link parameters is also a key aspect of cross-validatory posterior predictive methods, including ``node-splitting'', for assessing conflict across subsets of evidence \citep{Presanis:2013hb,Gasemyr:2009jt}.
Markov melding may provide a natural, computationally-efficient approach for systematic conflict assessment {\protect\citep{Presanis:2016}}.

Our framework can also be viewed as encapsulating a range of approaches proposed in the big data literature for handling a large number of observations (`tall data').
With tall data it may be infeasible even to store all of the data on a single computer, nevermind evaluate functions depending on the whole dataset thousands of times, as needed in MCMC.
Instead, a divide-and-conquer approach can be taken, in which the original exchangeable data \(\talldata\) are partitioned into \(\batchLast\) batches \(\talldata_{\batchFirst}, \dots, \talldata_{\batchLast}\), each of which contains few enough observations that standard statistical methods can be applied without undue trouble.
The key observation is that the full posterior distribution \(\p(\talltheta \given \talldata)\) can be split into a number of submodel posteriors \(\p_{\batchIndex}(\talltheta \given \talldata_{\batchIndex}) \propto p(\talldata_{\batchIndex} \given \talltheta)\p(\talltheta)^{1/\batchSize}, \quad \batchIndex = \batchFirst, \dots, \batchLast\).
This is a form of model splitting (Section~\ref{sec:splitting-with-markov-melding}), with \poe{} pooling and the original prior apportioned equally among the batches.
Various approaches for integrating the batch-specific posteriors to approximate the overall posterior have been proposed \citep{Huang:2005fo, Scott:2013wza, Neiswanger:2013ug, Wang:2013tg, Bardenet:2015vm, Minsker:2014wg}.
However, this literature has so far only considered independent, identically distributed data, whereas we have considered more general models and data.

\subsection{Computational challenges}
In our examples, the link variable \(\cpar\) is comparatively low dimensional and simple kernel density estimation using a multivariate $t$-distribution kernel proved sufficient.
Moreover, the results were robust with respect to the choice of kernel and kernel bandwidth.
For higher-dimensional link variables more care in the choice of kernel estimation method might be required \citep{henderson2015applied}, or, alternatively, we might wish to estimate the ratio of densities directly to improve stability \citep{sugiyama2012density}.

The multi-stage sampler (Section~\ref{sec:multi-stage-mwg}) broadly falls into the category of a sequential Monte Carlo sampler \citep{smith2013sequential}, as described in Supplementary Material~\ref{s:appendix-smc}.
While the Markov melding model is invariant to the ordering of the submodels used (assuming the pooling function is also), the efficiency of the multi-stage algorithm may not be in practice, due to the need for there to be sufficient stage one samples in the appropriate region.
If two submodels contain an approximately equal amount of non-conflicting information, then the ordering is unlikely to be important.
In other settings, more care may be required.
For example, suppose submodel \(M_{1}\) contains considerably more information than \(M_{2}\).
If stage one uses \(M_{1}\), then the stage one posterior may be so precise that it is unable to be adjusted for the extra information in \(M_{2}\).
In contrast, if \(M_{2}\) is used first, then the estimate of the posterior distribution may be very coarse, due to a lack of samples in the central part of the posterior distribution.
Further research will be needed to identify the best ordering to adopt in general.

\subsection{Alternative approaches}
We obtained consistency in the link parameter \(\cpar\), as required by Markov combination, through marginal replacement (Section~\ref{sec:joining-models}).
This approach assumes the priors differ in substance across submodels.
Alternatively, as we outline in Supplementary Material~\ref{s:copula}, we could assume that the priors differ only due to different scalings in each submodel, and so can be made  consistent through rescaling, similar to when deriving multivariate distributions from copulas \citep{durante2010copula}.
Yet another approach is a supra-Bayesian approach {\protect\citep{Lindley:1979vl,Leonelli:vn}}, in which the decision maker models the experts' opinions.

The prior pooling approach considered within our framework includes a judgement as to how to weight the different submodels.
Various other methods have been proposed for weighting evidence, including the cut operator \citep{Lunn:2013,Plummer2015}; the power prior approach in clinical trials \citep[for example, ][]{Neuenschwander:2009}; and modularisation in the computer models literature \citep{Liu:2009hm}.
Further research is required to investigate the relationship of Markov melding to other weighting approaches.

\section*{Acknowledgements}
\begin{sloppypar}
{\small This work was supported by the UK Medical Research Council [programme codes MC\_UP\_1302/3, MC\_U105260556, U105260557 and MC\_U105260799].
We are grateful to Ian White, Sylvia Richardson, Brian Tom, Michael Sweeting, Paul Kirk, Adrian Raftery, and the 2015 Armitage lecturers (Leonhard Held and Michael H\"{o}hle) for helpful discussions of this work.
We also thank colleagues at Public Health England for providing data.}
\end{sloppypar}

\bibliographystyle{../../bib/mystyle4}
\bibliography{paper}

\begin{thebibliography}{59}
\providecommand{\natexlab}[1]{#1}

\bibitem[{Ades and Sutton(2006)}]{Ades:2006tw}
Ades, A.~E. and Sutton, A.~J. (2006) {Multiparameter evidence synthesis in
  epidemiology and medical decision-making: current approaches}.
\newblock \emph{Journal of the Royal Statistical Society: Series A (Statistics
  in Society)}, \textbf{169}, 5--35.

\bibitem[{Albert \emph{et~al.}(2011)Albert, Espi{\'e}, de~Valk and
  Denis}]{Albert:2011em}
Albert, I., Espi{\'e}, E., de~Valk, H. and Denis, J.-B. (2011) {A Bayesian
  evidence synthesis for estimating Campylobacteriosis prevalence}.
\newblock \emph{Risk Analysis}, \textbf{31}, 1141--1155.

\bibitem[{Bardenet \emph{et~al.}(2017)Bardenet, Doucet and
  Holmes}]{Bardenet:2015vm}
Bardenet, R., Doucet, A. and Holmes, C. (2017) {On Markov chain Monte Carlo
  methods for tall data}.
\newblock \emph{Journal of Machine Learning Research}, \textbf{18}, 1--43.

\bibitem[{Besbeas \emph{et~al.}(2002)Besbeas, Freeman, Morgan and
  Catchpole}]{Besbeas:2002jw}
Besbeas, P., Freeman, S.~N., Morgan, B. J.~T. and Catchpole, E.~A. (2002)
  {Integrating mark{\textendash}recapture{\textendash}recovery and census data
  to estimate animal abundance and demographic parameters}.
\newblock \emph{Biometrics}, \textbf{58}, 540--547.

\bibitem[{{Birrell} \emph{et~al.}(2016){Birrell}, {De Angelis}, {Wernisch},
  {Tom}, {Roberts} and {Pebody}}]{Birrell2016}
{Birrell}, P.~J., {De Angelis}, D., {Wernisch}, L., {Tom}, B.~D., {Roberts},
  G.~O. and {Pebody}, R.~G. (2016) {Efficient real-time monitoring of an
  emerging influenza epidemic: how feasible?}
\newblock {arXiv:1608.05292}.

\bibitem[{Brooks \emph{et~al.}(2004)Brooks, King and Morgan}]{Brooks:2004wr}
Brooks, S.~P., King, R. and Morgan, B. J.~T. (2004) {A Bayesian approach to
  combining animal abundance and demographic data}.
\newblock \emph{Animal Biodiversity and Conservation}, \textbf{27}, 515--529.

\bibitem[{Clemen and Winkler(1999)}]{Clemen:1999vy}
Clemen, R.~T. and Winkler, R.~L. (1999) {Combining probability distributions
  from experts in risk analysis}.
\newblock \emph{Risk Analysis}, \textbf{19}, 187--203.

\bibitem[{Commenges and Hejblum(2012)}]{Commenges:2012gv}
Commenges, D. and Hejblum, B.~P. (2012) {Evidence synthesis through a
  degradation model applied to myocardial infarction}.
\newblock \emph{Lifetime Data Analysis}, \textbf{19}, 1--18.

\bibitem[{Dawid and Lauritzen(1993)}]{Dawid:1993fc}
Dawid, A.~P. and Lauritzen, S.~L. (1993) {Hyper Markov laws in the statistical
  analysis of decomposable graphical models}.
\newblock \emph{Annals of Statistics}, \textbf{21}, 1272--1317.

\bibitem[{{Department of Health}(2011)}]{DoHWW:2011}
{Department of Health} (2011) {Department of Health Winter Watch}.
\newblock {http://winterwatch.dh.gov.uk}.

\bibitem[{Doucet \emph{et~al.}(2013)Doucet, de~Freitas and
  Gordon}]{smith2013sequential}
Doucet, A., de~Freitas, N. and Gordon, N., eds. (2013) \emph{{Sequential Monte
  Carlo Methods in Practice}}.
\newblock New York: Springer Science \& Business Media.

\bibitem[{Draper(1995)}]{Draper:1995ts}
Draper, D. (1995) {Assessment and propagation of model uncertainty}.
\newblock \emph{Journal of the Royal Statistical Society: Series B
  (Methodological)}, \textbf{57}, 45--97.

\bibitem[{Durante and Sempi(2010)}]{durante2010copula}
Durante, F. and Sempi, C. (2010) Copula theory: an introduction.
\newblock In \emph{Copula Theory and its Applications} (eds. P.~Jaworski,
  F.~Durante, W.~K. H\"{a}rdle and T.~Rychlik), pp. 3--31. Berlin:
  Springer-Verlag.

\bibitem[{Eddy \emph{et~al.}(1992)Eddy, Hasselblad and Shachter}]{EddyEtAl1992}
Eddy, D.~M., Hasselblad, V. and Shachter, R. (1992) \emph{{Meta-Analysis} by
  the Confidence Profile Method}.
\newblock London: Academic Press.

\bibitem[{G{\aa}semyr and Natvig(2009)}]{Gasemyr:2009jt}
G{\aa}semyr, J. and Natvig, B. (2009) {Extensions of a conflict measure of
  inconsistencies in Bayesian hierarchical models}.
\newblock \emph{Scandinavian Journal of Statistics}, \textbf{36}, 822--838.

\bibitem[{Genest and Zidek(1986)}]{Genest:1986be}
Genest, C. and Zidek, J.~V. (1986) {Combining probability distributions: A
  critique and an annotated bibliography}.
\newblock \emph{Statistical Science}, \textbf{1}, 114--135.

\bibitem[{Gilks and Berzuini(2001)}]{gilks2001following}
Gilks, W.~R. and Berzuini, C. (2001) Following a moving target---Monte Carlo
  inference for dynamic Bayesian models.
\newblock \emph{Journal of the Royal Statistical Society: Series B (Statistical
  Methodology)}, \textbf{63}, 127--146.

\bibitem[{Goudie \emph{et~al.}(2015)Goudie, Hovorka, Murphy and
  Lunn}]{Goudie:2015kg}
Goudie, R. J.~B., Hovorka, R., Murphy, H.~R. and Lunn, D. (2015) {Rapid model
  exploration for complex hierarchical data: application to pharmacokinetics of
  insulin aspart.}
\newblock \emph{Statistics in Medicine}, \textbf{34}, 3144--3158.

\bibitem[{Green \emph{et~al.}(2003)Green, Hjort and
  Richardson}]{hssIntroduction2003}
Green, P.~J., Hjort, N.~L. and Richardson, S. (2003) {Introducing highly
  structured stochastic systems}.
\newblock In \emph{Highly Structured Stochastic Systems} (eds. P.~J. Green,
  N.~L. Hjort and S.~Richardson), pp. 1--12. Oxford: Oxford University Press.

\bibitem[{Hasselblad \emph{et~al.}(1992)Hasselblad, Eddy and
  Kotchmar}]{Hasselblad:1992iw}
Hasselblad, V., Eddy, D.~M. and Kotchmar, D.~J. (1992) {Synthesis of
  environmental evidence: nitrogen dioxide epidemiology studies}.
\newblock \emph{Journal of the Air \& Waste Management Association},
  \textbf{42}, 662--671.

\bibitem[{Henderson and Parmeter(2015)}]{henderson2015applied}
Henderson, D.~J. and Parmeter, C.~F. (2015) \emph{{Applied Nonparametric
  Econometrics}}.
\newblock New York: Cambridge University Press.

\bibitem[{Hinton(2002)}]{Hinton:2002ic}
Hinton, G.~E. (2002) {Training products of experts by minimizing contrastive
  divergence.}
\newblock \emph{Neural computation}, \textbf{14}, 1771--1800.

\bibitem[{Huang and Gelman(2005)}]{Huang:2005fo}
Huang, Z. and Gelman, A. (2005) {Sampling for Bayesian computation with large
  datasets}.
\newblock Working paper.

\bibitem[{Jackson \emph{et~al.}(2009)Jackson, Best and
  Richardson}]{Jackson:2009km}
Jackson, C.~H., Best, N.~G. and Richardson, S. (2009) {Bayesian graphical
  models for regression on multiple data sets with different variables.}
\newblock \emph{Biostatistics}, \textbf{10}, 335--351.

\bibitem[{Jackson \emph{et~al.}(2015)Jackson, Jit, Sharples and
  Angelis}]{Jackson2015}
Jackson, C.~H., Jit, M., Sharples, L.~D. and Angelis, D.~D. (2015) Calibration
  of complex models through Bayesian evidence synthesis.
\newblock \emph{Medical Decision Making}, \textbf{35}, 148--161.

\bibitem[{Lauritzen(1996)}]{Lauritzen96}
Lauritzen, S.~L. (1996) \emph{Graphical Models}.
\newblock Oxford: Clarendon Press.

\bibitem[{Leonelli(2015)}]{Leonelli:vn}
Leonelli, M. (2015) \emph{Bayesian decision support in complex modular systems:
  an algebraic and graphical approach}.
\newblock Ph.D. thesis, University of Warwick, UK.

\bibitem[{Liang and Weiss(2007)}]{Liang:2007fi}
Liang, L.-J. and Weiss, R.~E. (2007) {A hierarchical semiparametric regression
  model for combining HIV-1 phylogenetic analyses using iterative reweighting
  algorithms}.
\newblock \emph{Biometrics}, \textbf{63}, 733--741.

\bibitem[{Lindley \emph{et~al.}(1979)Lindley, Tversky and
  Brown}]{Lindley:1979vl}
Lindley, D.~V., Tversky, A. and Brown, R.~V. (1979) {On the reconciliation of
  probability assessments}.
\newblock \emph{Journal of the Royal Statistical Society: Series A (General)},
  \textbf{142}, 146--180.

\bibitem[{Liu \emph{et~al.}(2009)Liu, Bayarri and Berger}]{Liu:2009hm}
Liu, F., Bayarri, M.~J. and Berger, J.~O. (2009) {Modularization in Bayesian
  analysis, with emphasis on analysis of computer models}.
\newblock \emph{Bayesian Analysis}, \textbf{4}, 119--150.

\bibitem[{Liu and West(2001)}]{liu2001combined}
Liu, J. and West, M. (2001) \emph{Combined parameter and state estimation in
  simulation-based filtering}, pp. 197--223.
\newblock In  \cite{smith2013sequential}.

\bibitem[{Lunn \emph{et~al.}(2013{\natexlab{a}})Lunn, Barrett, Sweeting and
  Thompson}]{Lunn:2013uo}
Lunn, D., Barrett, J., Sweeting, M. and Thompson, S. (2013{\natexlab{a}})
  {Fully Bayesian hierarchical modelling in two stages, with application to
  meta-analysis.}
\newblock \emph{Journal of the Royal Statistical Society: Series C (Applied
  Statistics)}, \textbf{62}, 551--572.

\bibitem[{Lunn \emph{et~al.}(2013{\natexlab{b}})Lunn, Jackson, Best, Thomas and
  Spiegelhalter}]{Lunn:2013}
Lunn, D., Jackson, C., Best, N., Thomas, A. and Spiegelhalter, D.
  (2013{\natexlab{b}}) \emph{The BUGS Book: A Practical Introduction to
  Bayesian Analysis}.
\newblock Boca Raton: CRC Press.

\bibitem[{Massa and Lauritzen(2010)}]{Massa:2010uf}
Massa, M.~S. and Lauritzen, S.~L. (2010) {Combining statistical models}.
\newblock In \emph{Contemporary Mathematics: Algebraic Methods in Statistics
  and Probability II} (eds. M.~A.~G. Viana and H.~P. Wynn), pp. 239--260.

\bibitem[{Massa and Riccomagno(2017)}]{Massa:2017bz}
Massa, M.~S. and Riccomagno, E. (2017) {Algebraic representations of Gaussian
  Markov combinations}.
\newblock \emph{Bernoulli}, \textbf{23}, 626--644.

\bibitem[{Miller and Dunson(2015)}]{Miller:2015tq}
Miller, J.~W. and Dunson, D.~B. (2015) {Robust Bayesian inference via
  coarsening}.
\newblock {arXiv:1506.06101}.

\bibitem[{Minsker \emph{et~al.}(2014)Minsker, Srivastava, Lin and
  Dunson}]{Minsker:2014wg}
Minsker, S., Srivastava, S., Lin, L. and Dunson, D.~B. (2014) {Robust and
  scalable Bayes via a median of subset posterior measures}.
\newblock {arXiv:1403.2660v3}.

\bibitem[{Moran and Clark(2011)}]{MoranClark2011}
Moran, E.~V. and Clark, J.~S. (2011) Estimating seed and pollen movement in a
  monoecious plant: a hierarchical Bayesian approach integrating genetic and
  ecological data.
\newblock \emph{Molecular Ecology}, \textbf{20}, 1248--1262.

\bibitem[{M{\"u}ller(1991)}]{Muller:1991wq}
M{\"u}ller, P. (1991) {A generic approach to posterior integration and Gibbs
  sampling}.
\newblock Technical Report 91-09, Purdue University.

\bibitem[{Neiswanger \emph{et~al.}(2014)Neiswanger, Wang and
  Xing}]{Neiswanger:2013ug}
Neiswanger, W., Wang, C. and Xing, E.~P. (2014) Asymptotically exact,
  embarrassingly parallel MCMC.
\newblock In \emph{Proceedings of the Thirtieth Conference Annual Conference on
  Uncertainty in Artificial Intelligence (UAI-14)}, pp. 623--632. Corvallis,
  Oregon: AUAI Press.

\bibitem[{Neuenschwander \emph{et~al.}(2009)Neuenschwander, Branson and
  Spiegelhalter}]{Neuenschwander:2009}
Neuenschwander, B., Branson, M. and Spiegelhalter, D.~J. (2009) A note on the
  power prior.
\newblock \emph{Statistics in Medicine}, \textbf{28}, 3562--3566.

\bibitem[{O'Hagan \emph{et~al.}(2006)O'Hagan, Buck, Daneshkhah, Eiser,
  Garthwaite, Jenkinson, Oakley and Rakow}]{OHagan:2006}
O'Hagan, A., Buck, C.~E., Daneshkhah, A., Eiser, J.~R., Garthwaite, P.~H.,
  Jenkinson, D.~J., Oakley, J.~E. and Rakow, T. (2006) \emph{Uncertain
  Judgements: Eliciting Experts' Probabilities}.
\newblock Chichester: John Wiley \& Sons.

\bibitem[{Plummer(2015{\natexlab{a}})}]{Plummer2015}
Plummer, M. (2015{\natexlab{a}}) Cuts in Bayesian graphical models.
\newblock \emph{Statistics and Computing}, \textbf{25}, 37--43.

\bibitem[{Plummer(2015{\natexlab{b}})}]{plummer2015jags}
Plummer, M. (2015{\natexlab{b}}) JAGS Version 4.0.1 user manual.

\bibitem[{Poole and Raftery(2000)}]{Poole:2000ee}
Poole, D. and Raftery, A.~E. (2000) {Inference for deterministic simulation
  models: The Bayesian melding approach}.
\newblock \emph{Journal of the American Statistical Association}, \textbf{95},
  1244--1255.

\bibitem[{Presanis \emph{et~al.}(2016)Presanis, Ohlssen, Cui, Rosinska and
  De~Angelis}]{Presanis:2016}
Presanis, A.~M., Ohlssen, D., Cui, K., Rosinska, M. and De~Angelis, D. (2016)
  Conflict diagnostics for evidence synthesis in a multiple testing framework.
\newblock {arXiv:1702.07304}.

\bibitem[{Presanis \emph{et~al.}(2013)Presanis, Ohlssen, Spiegelhalter and
  De~Angelis}]{Presanis:2013hb}
Presanis, A.~M., Ohlssen, D., Spiegelhalter, D.~J. and De~Angelis, D. (2013)
  {Conflict diagnostics in directed acyclic graphs, with applications in
  Bayesian evidence synthesis}.
\newblock \emph{Statistical Science}, \textbf{28}, 376--397.

\bibitem[{Presanis \emph{et~al.}(2014)Presanis, Pebody, Birrell, Tom, Green,
  Durnall, Fleming and De~Angelis}]{Presanis:2014em}
Presanis, A.~M., Pebody, R.~G., Birrell, P.~J., Tom, B. D.~M., Green, H.~K.,
  Durnall, H., Fleming, D. and De~Angelis, D. (2014) {Synthesising evidence to
  estimate pandemic (2009) A/H1N1 influenza severity in 2009{\textendash}2011}.
\newblock \emph{Annals of Applied Statistics}, \textbf{8}, 2378--2403.

\bibitem[{{Public Health England}(2014)}]{PHEdataMart:2014}
{Public Health England} (2014) {Sources of UK flu data: influenza surveillance
  in the UK}.
\newblock
  {https://www.gov.uk/guidance/sources-of-uk-flu-data-influenza-surveillance-in-the-uk}.

\bibitem[{Robert and Casella(2004)}]{robertCasellaMCMC}
Robert, C.~P. and Casella, G. (2004) \emph{{Monte Carlo Statistical Methods}}.
\newblock New York: Springer.

\bibitem[{Scott \emph{et~al.}(2016)Scott, Blocker, Bonassi, Chipman, George and
  McCulloch}]{Scott:2013wza}
Scott, S.~L., Blocker, A.~W., Bonassi, F.~V., Chipman, H.~A., George, E.~I. and
  McCulloch, R.~E. (2016) {Bayes and big data: The consensus Monte Carlo
  algorithm}.
\newblock \emph{International Journal of Management Science and Engineering
  Management}, \textbf{11}, 78--88.

\bibitem[{Shubin \emph{et~al.}(2016)Shubin, Lebedev, Lyytik\"{a}inen and
  Auranen}]{Shubin2016}
Shubin, M., Lebedev, A., Lyytik\"{a}inen, O. and Auranen, K. (2016) Revealing
  the true incidence of pandemic A(H1N1)pdm09 influenza in Finland during the
  first two seasons---An analysis based on a dynamic transmission model.
\newblock \emph{PLOS Computational Biology}, \textbf{12}, e1004803.

\bibitem[{Sugiyama \emph{et~al.}(2012)Sugiyama, Suzuki and
  Kanamori}]{sugiyama2012density}
Sugiyama, M., Suzuki, T. and Kanamori, T. (2012) \emph{{Density Ratio
  Estimation in Machine Learning}}.
\newblock New York: Cambridge University Press.

\bibitem[{Tom \emph{et~al.}(2010)Tom, Sinsheimer and Suchard}]{Tom:2010uf}
Tom, J.~A., Sinsheimer, J.~S. and Suchard, M.~A. (2010) {Reuse, recycle,
  reweigh: combating influenza through efficient sequential Bayesian
  computation for massive data}.
\newblock \emph{Annals of Applied Statistics}, \textbf{4}, 1722--1748.

\bibitem[{Turner \emph{et~al.}(2009)Turner, Spiegelhalter, Smith and
  Thompson}]{Turner:2009}
Turner, R.~M., Spiegelhalter, D.~J., Smith, G. C.~S. and Thompson, S.~G. (2009)
  Bias modelling in evidence synthesis.
\newblock \emph{Journal of the Royal Statistical Society: Series A (Statistics
  in Society)}, \textbf{172}, 21--47.

\bibitem[{Wang and Dunson(2013)}]{Wang:2013tg}
Wang, X. and Dunson, D.~B. (2013) {Parallel MCMC via Weierstrass Sampler}.
\newblock {arXiv:1312.4605}.

\bibitem[{Welton \emph{et~al.}(2008)Welton, Cooper, Ades, Lu and
  Sutton}]{Welton:2008iy}
Welton, N.~J., Cooper, N.~J., Ades, A.~E., Lu, G. and Sutton, A.~J. (2008)
  {Mixed treatment comparison with multiple outcomes reported inconsistently
  across trials: Evaluation of antivirals for treatment of influenza A and B}.
\newblock \emph{Statistics in Medicine}, \textbf{27}, 5620--5639.

\bibitem[{Welton \emph{et~al.}(2012)Welton, Sutton, Cooper, Abrams and
  Ades.}]{Welton:2012}
Welton, N.~J., Sutton, A.~J., Cooper, N.~J., Abrams, K.~R. and Ades., A. (2012)
  \emph{{Evidence Synthesis for Decision Making in Healthcare}}.
\newblock Chichester: John Wiley \& Sons.

\bibitem[{Wilkinson(2013)}]{Wilkinson:2013ci}
Wilkinson, R.~D. (2013) {Approximate Bayesian computation (ABC) gives exact
  results under the assumption of model error.}
\newblock \emph{Statistical Applications in Genetics and Molecular Biology},
  \textbf{12}, 129--141.

\end{thebibliography}

\section*{Supplementary Material}

\setcounter{section}{0}
\setcounter{equation}{0}
\setcounter{table}{0}
\setcounter{figure}{0}
\renewcommand{\theequation}{S\arabic{equation}}
\renewcommand{\thesection}{\Alph{section}}
\renewcommand{\thetable}{S\arabic{table}}
\renewcommand{\thefigure}{S\arabic{figure}}

\section{Motivation for marginal replacement}
\label{s:appendix-kl}

We argue that a motivation for~(\ref{e:marginal-replacement}) is that $\prepl$ minimises the Kullback-Leibler (KL) divergence of a distribution $q(\phi,\psi_m,Y_m)$ to $p(\phi,\psi_m,Y_m)$ under the constraint that the marginals on $\phi$ agree, $q(\phi) = \pnew(\phi)$, that is,
\[ \prepl(\phi,\psi_m,Y_m) = \mbox{argmin}_q
  \{ D_\mathrm{KL}(q \parallel p_m) \mid
            q(\phi) = \pnew(\phi) \mbox{ for all $\phi$} \} \]
This is easily shown as follows (we drop index $m$ and variable $Y$ for simplicity).
The KL divergence under the constraint is given by
\begin{equation*}\begin{split}
 D_\mathrm{KL}(q \parallel p) &=
 \int q(\phi,\psi) \log\frac{q(\phi,\psi)}{p(\phi,\psi)}\,d\phi\,d\psi \\
 &=  \int q(\psi\mid \phi)
       \log\frac{q(\psi\mid\phi)}{p(\psi\mid\phi)} \,d\psi\,q(\phi)\,d\phi +
    \int q(\phi,\psi)
       \,d\psi \log\frac{q(\phi)}{p(\phi)} \,d\phi \\
 &=  \int D_\mathrm{KL}(q(\cdot\mid \phi) \parallel p(\cdot\mid \phi))\, q(\phi)\,d\phi +
    \int \pnew(\phi) \log\frac{\pnew(\phi)}{p(\phi)} \,d\phi
\end{split}\end{equation*}
The second term is the KL divergence of the marginals and is constant.
The first term can be minimised to 0 by choosing $q(\psi\mid\phi)= p(\psi \mid \phi)$ for all $\phi$ and consequently $q(\phi,\psi) = p(\psi \mid \phi)\pnew(\phi)$ is the solution to the constrained KL divergence minimisation.
A similar argument has been made in~\citet{Poole:2000ee} to justify their choice of a distribution for Bayesian melding.
Notice that the same argument can still be made based on $D_\mathrm{KL}(p_m \parallel q)$ with the roles of $p_m$ and $q$ exchanged.

Marginal replacement can also be seen as a generalisation of Bayesian updating in the light of new information.
For example, if we learn that $\phi$ can assume only the value of a constant $\phi_0$, standard Bayesian updating entails conditioning on the new information $\phi_0$ to form the posterior distribution $p_m(\psi_m,Y_m\mid \phi_0)$.
This can be viewed as a special case of marginal replacement in which the new marginal $\pnew(\phi)$ is the point mass \(\delta_{\phi_0}(\phi)\) on $\phi_0$, since the marginal distribution of \((\psi_m, Y_m)\) under the marginal replacement model is this posterior distribution, as follows by standard properties of the Dirac delta function~$\delta_{\phi_0}$:
\begin{equation}\begin{split}\label{e:marginal-replacement-delta}
\preplm(\psi_m,Y_m)
&= \int \preplm(\phi,\psi_m,Y_m)\, d\phi\\
&=
\int p_m(\psi_m,Y_m \mid \phi)\, \delta_{\phi_0}(\phi)\,d\phi
= p_m(\psi_m,Y_m\mid \phi_0)
\end{split}\end{equation}
 In this sense, (\ref{e:marginal-replacement}) enables integration of new information on $\phi$ provided not only in the form of a specific value $\phi_0$ but in the form of a general density function $\pnew(\phi)$.

Finally, Approximate Bayesian Computation (ABC) can be interpreted as a marginal replacement similar to the replacement in equation~(\ref{e:marginal-replacement-delta}), when $\phi$ typically represents a data variable\footnote{We thank Paul Kirk for this observation}.
Instead of $\pnew(\phi) = \delta_{\phi_0}(\phi)$ in standard posterior inference, ABC uses
\[ \pnew(\phi)\allowbreak=p(\phi)\,\allowbreak I(d(S(\phi),S(\phi_0)) < \epsilon) \]
 where $I$ is the indicator function of an event, $d$ is a distance function, $S$ some summary statistic for the data variable $\phi$, $\phi_0$ is observed and $\epsilon$ a small constant.
ABC can thus be seen as very similar to standard posterior inference but with a widening of the $\delta$ function \citep{Wilkinson:2013ci, Miller:2015tq}.
In fact, the limits $\epsilon \to 0$ and $\epsilon \to \infty$ lead to the posterior and prior distributions on $\phi$, respectively.

\section{Transformations with noninvertible deterministic functions}
\label{s:appendix-deterministic}

$\theta$ is a $\ell$-dimensional real multivariate variable, and $\phi(\theta)=(\phi_1(\theta),\ldots,\phi_k(\theta))$, $k < \ell$, a deterministic transformation that can be expanded to an invertible function $\phi_e(\theta) = (\phi(\theta),t(\theta))$, for $t(\theta) = (t_1(\theta),\ldots,t_{\ell-k}(\theta))$.
We assume the inverse mapping $\theta(\phi,t)$ and $\phi_e(\theta)$ have first derivatives.
Mapping $\phi_e$ induces a probability distribution on $(\phi,t)$ which can be represented as
\begin{equation}\label{e:jacobian-app}
 p(\phi,t) = p(\theta(\phi,t))\, J_\theta(\phi,t)
\end{equation}
 where $J_\theta(\phi,t)$ is the Jacobian determinant for the transformation $\theta(\phi,t)$.
The induced marginal distribution on $\phi$ can then be defined as
 \begin{equation}\label{e:marginal-jacobian-app}
  p(\phi) = \int p(\phi,t) \, dt
 \end{equation}
Recall that the Jacobian determinant of the inverse transformation $\theta(\phi,t)$ is
 \[ J_\theta(\phi,t)
 = \begin{vmatrix}
    {\partial \theta}/{\partial \phi} \\ {\partial \theta}/{\partial t}
    \end{vmatrix}_{(\phi,t)}
 = (J_{(\phi,t)}(\theta))^{-1} = \left |\frac{\partial (\phi,t)}{\partial \theta} \right|_{\theta(\phi,t)}^{-1}
 = \begin{vmatrix}
    {\partial \phi}/{\partial \theta} \\ {\partial t}/{\partial \theta}
    \end{vmatrix}_{\theta(\phi,t)}^{-1}
\]
 where any ${\partial u}/{\partial v} = (\partial u_i/\partial v_j)_{ij}$ is the matrix of partial derivatives of functions~$u_i$ by variables~$v_j$ and $|\cdot|$ denotes the absolute value of the determinant.

Here we show that the value of $p(\phi)$ is independent of the particular parameterisation.
That is, if $s(\theta)$ is an alternative parameterisation so that $\widetilde{\phi}_e(\theta) = (\phi(\theta),s(\theta))$ also has an inverse mapping $\theta(\phi,s)$ then for a {\em fixed\/} $\phi$ we have an invertible transformation $s(t) = s(\theta(\phi,t))$ and
\begin{equation*}\begin{split}
  \int p(\phi,s) \, ds &= \int p(\phi,s(t)) \frac{ds}{dt}(t)\, dt
  = \int p(\theta(\phi,s(t)))
    \begin{vmatrix}
    {\partial \theta}/{\partial \phi} \\ {\partial \theta}/{\partial s}
    \end{vmatrix}_{\theta(\phi,s(t))} \frac{ds}{dt}(t) \, dt \\
  &= \int p(\theta(\phi,t))
    \begin{vmatrix}
    {\partial \theta}/{\partial \phi} \\
               {\partial \theta}/{\partial s}\,ds/dt
    \end{vmatrix}_{\theta(\phi,t)} \, dt
    = \int p(\theta(\phi,t))
    \begin{vmatrix}
    {\partial \theta}/{\partial \phi} \\ {\partial \theta}/{\partial t}
    \end{vmatrix}_{\theta(\phi,t)} \, dt  \\
  &= \int p(\phi,t) \, dt = p(\phi)\\
\end{split}\end{equation*}
where we used the multilinearity of the determinant, the chain rule for multidimensional derivatives, and that $\theta(\phi,s(t)) = \theta(\phi,t)$ by the definition of $s(t)$.
Consequently, $p(\phi)$ as the induced probability on all values taken by $\phi(\theta)$ is well defined.

The marginal distribution~(\ref{e:marginal-jacobian-app}) of \(\cpar\) in the density $p(\phi,t)$  in~(\ref{e:jacobian-app}) can then be replaced by any other desired marginal $\pnew(\phi)$ via marginal replacement as in~(\ref{e:marginal-replacement})
\begin{equation*}%
 \prepl(\phi,t) = \frac{p(\phi,t)}{p(\phi)}\pnew(\phi) =  \frac{p(\theta(\phi,t)) J_\theta(\phi,t)}{p(\phi)} \pnew(\phi)
\end{equation*}
Finally, the new distribution $\prepl(\phi,t)$  is mapped back to $\theta$ using the invertible mapping $\phi_e(\theta)=(\phi(\theta),t(\theta))$
\begin{equation}\label{e:raftery-equivalent}
 \prepl(\theta) = \prepl(\phi(\theta),t(\theta)) J_\theta(\phi,t)^{-1} = p(\theta) \frac{\pnew(\phi(\theta))}{p(\phi(\theta))}
\end{equation}
 which results in equation~(\ref{e:marginal-replacement-deterministic}).

 The last equation is similar to equation~(16) in \citet{Poole:2000ee}.
One of the key issues in their study is how to distribute the probability density at $\pnew(\phi_0)$ over $\theta$ with $\phi(\theta) = \phi_0$.
Equation~(16) in
\citet{Poole:2000ee} as well as equation~(\ref{e:raftery-equivalent}) here suggest doing this in proportion $p(\theta)/p(\phi(\theta))$ of the contribution of density $p(\theta)$ to $p(\phi(\theta)) = p(\phi_0)$.  \citet{Poole:2000ee} justify this approach more directly by using Kullback-Leibler divergence similar to Supplementary Material~\ref{s:appendix-kl} above.
Here it is a consequence of our slightly more general marginal replacement framework, which can also be justified by a Kullback-Leibler divergence argument as in Supplementary Material~\ref{s:appendix-kl}.

\section{Externally Bayesian pooling}
\label{s:externally-bayesian-pooling}
When it comes to the choice of a pooling strategy, one might want to consider the following argument for logarithmic pooling. A pooling strategy $g(p_1(\phi),\ldots,p_M(\phi))$ for the priors of $M$ distributions $p_i(\phi,Y)=p(Y\mid \phi)p_i(\phi)$ is called \emph{externally Bayesian} (EB) if it also applies to the posteriors in the sense that $g(p_1(\phi\mid Y),\ldots,p_M(\phi\mid Y) \propto p(Y\mid \phi)g(p_1(\phi),\ldots,p_M(\phi))$, that is, Bayesian updating and pooling are interchangeable. In this sense logarithmic pooling with $\sum w_i = 1$ is EB, which has been used to argue for its superiority over other pooling functions \citep{Genest:1986be}. However, EB is not applicable when combining several likelihoods with distinct data, since it is not the case that $g(p_1(\phi,\psi_1\mid Y_1),\ldots,p_M(\phi,\psi_M\mid Y_M) \propto \prod_i p_i(Y_i,\psi_i\mid \phi)g(p_1(\phi),\ldots,p_M(\phi))$. In a limited sense the EB property is relevant in the melding context if we wish to compromise between $M$ submodels $p_m(Y,\psi,\phi)$ which all use the same likelihood $p(Y\mid \phi) = \prod_i p_i(Y_i,\psi_i\mid \phi)$, but different priors $p_m(\phi)$, that is, $p_m(Y,\psi,\phi)=\prod_i p_i(Y_i,\psi_i\mid \phi)\,p_m(\phi)$, $m=1,\ldots,M$, ($M$ melding distributions with different dictatorial poolings). In this sense melding with log pooling is an EB compromise for $M$ individual meldings with dictatorial pooling.

\section{Multi-stage and sequential Monte Carlo sampling}
\label{s:appendix-smc}

A high-level feature of a sequential Monte Carlo (SMC) approach \citep{smith2013sequential} is the aim to obtain a sample~$S_M$ from a distribution~$\pi_M$ via sampling from intermediate distributions $\pi_1,\ldots,\pi_M$ producing samples $S_1,\ldots,S_M$, where~$S_{\stage-1}$ is used to produce~$S_\stage$.
In this broad sense our multi-stage sampler is an example of such an algorithm.
There are, however, deviations from a typical implementation of an SMC approach.

Formally, our target distributions are
 \[ \pi_\stage(\phi,\psi_1,\ldots,\psi_\stage) \propto \p_{\mathrm{meld},\stage}(\phi,\psiallpart,\yallpart)
 = \prod_{m=1}^\stage \rho_m(\phi,\psi_m) \]
as in~(\ref{e:iterative-melded}).
For simplicity we assume we are mostly interested in tracking samples of $\phi$ through the stages: $S_\stage = \{\phi_\stage^{(1)},\ldots,\phi_\stage^{(n_M)}\}$.
If parameters $\psi_m$ can be marginalised over, one could employ a typical sequential importance sampling scheme: sample $S_\stage$ from $S_{\stage-1}$ with probability proportional to
 \[ w_\stage^{(i)} = \frac{\pi_\stage(\phi_{\stage-1}^{(i)})}{\pi_{\stage-1}(\phi_{\stage-1}^{(i)})}
  = \rho_\stage(\phi_{\stage-1}^{(i)}) \]
Note that we only need to evaluate the likelihood $\rho_m(\phi)$ for the last submodel $m=\stage$ due to the factorisation of $\pmeld$.
Equivalently, a sample can be obtained via Metropolis-Hastings sampling with target-to-proposal density ratio
 \[ R(\phi^\star,\phi) = \pi_\stage(\phi^\star) \times \frac{1}{q(\phi^\star)}
 = \pi_\stage(\phi^\star) \times \frac{1}{\pi_{\stage-1}(\phi^\star)} = \rho_\stage(\phi^\star) \]
 where the proposal functions $q(\phi^\star)$ just samples uniformly from $S_{\stage-1}$.

We opted for the latter sampling approach since for the models envisaged it is rarely possible to marginalise out $\psi_m$ and the Metropolis-Hastings sampler is able to sample from both $\phi$ and $\psi_m$ together.

A notorious problem with static parameters such as $\phi$ is depletion of the sample with fewer distinct values of $\phi$ at each stage.
Various schemes have been proposed to rejuvenate the sample.
\citet{liu2001combined} propose adding a disturbance $\zeta$ to $\phi$ at each stage.
This amounts to sampling from a kernel smoothed version of the original sample.
Care needs be taken to avoid undue increase in variance from stage to stage.
\citet{liu2001combined} show how the increase can be controlled by cleverly correlating the disturbance $\zeta$ with $\phi$.

\citet{gilks2001following} propose a rejuvenation through a move step after the sampling step.
This move step, applied at stage~$\stage$, needs to leave distribution $\pi_\stage$ invariant, for example, by one or more Metropolis-Hastings steps.
In our case this is only possible by evaluating the full distribution $p_{\mathrm{meld},\stage}$ involving all submodels $m=1,\ldots,\stage$, somehow defeating the purpose of the scheme to avoid revisiting submodels earlier than $\stage$.
However, for a long sequence of submodels an occasional move step might be beneficial despite the increase in computation.

\section{Transformation of marginals of the link variable}
\label{s:copula}

An alternative approach to achieve the same distribution of the link variable $\phi$ in all submodels required by~(\ref{e:markov-combination}) is via a suitable transformation of $\phi$ so that all marginals agree similar to a copula approach \citep{durante2010copula}.
We assume we have link variables $\phi_m$ for each submodel~$m$, measuring the same quantity (for example weight) but on different scales (say, kilograms, stones, pounds), which we indicate by a submodel-specific index~$m$ of~$\phi_m$.
However, the twist is that we cannot assume the transformations between scales are known.
Instead we assume they can be reconstructed by matching quantiles of the prior distributions on the link variables.
That is, find transformations so that all distributions are identical after rescaling.

We further assume we have a presentation of the link variable $\phi$ on a standard scale with distribution $\pnew(\phi)$.
For a suitable transformation for submodel~$m$, let $F_m(\phi_m)$ and $\fnew(\phi)$ denote the cumulative distribution function for $p_m(\phi_m)$ and $\pnew(\phi)$ and let $F_m^{-}(\phi)$ and $\fnew^{-}(\phi)$ denote their inverse functions.
The submodel-specific mappings $\phi_m=\transfun_{m,\rnew}(\phi)=F_m^{-}(\fnew(\phi))$ transform between $\phi_m$ and $\phi$ preserving their densities $p_m(\phi_m)$ and $\pnew(\phi)$:
 \begin{equation}\begin{split}\label{e:transform_m}
 p_m(\phi_m) = p_m(\transfun_{m,\rnew}(\phi))\frac{d\transfun_{m,\rnew}}{d\phi}(\phi)
=p_m(\transfun_{m,\rnew}(\phi))\frac{\pnew(\phi)}{p_m(\transfun_{m,\rnew}(\phi))} =
\pnew(\phi)
\end{split}\end{equation}

We are now able to define new distributions that agree in their marginals on~$\phi_m$ and~$\phi$ by applying transformations $\phi_m=\transfun_{m,\rnew}(\phi)$ to $p_m(\phi_m,\psi_m,Y_m)$
\begin{equation*}\begin{split}
p_{\rtransf,m}(\phi,\psi_m,Y_m)
 &= p_{m}(\transfun_{m,\rnew}(\phi),\psi_m,Y_m)\frac{d\transfun_{m,\rnew}}{d\phi}(\phi)\\
&=
\frac{p_m(\transfun_{m,\rnew}(\phi),\psi_m,Y_m)}{p_m(\transfun_{m,\rnew}(\phi))}\pnew(\phi)\\
&=
p_m(\psi_m,Y_m \mid \transfun_{m,\rnew}(\phi))\, \pnew(\phi)\\
\end{split}\end{equation*}
Applying Markov combination to these transformed submodels results in a joint distribution
\begin{equation}\begin{split}
\label{e:copula-melding}
\ptmeld(\phi,\psi_1,\ldots,\psi_M,Y_1,\ldots,Y_M)
 &= \pnew(\phi) \prod_{m=1}^M p_m(\psi_m,Y_m \mid \transfun_{m,\rnew}(\phi))\\
 &=\pnew(\phi) \prod_{m=1}^M \frac{p_m(\transfun_{m,\rnew}(\phi),\psi_m,Y_m)}{p_m(\transfun_{m,\rnew}(\phi))}
\end{split}\end{equation}

Remarkably, it is straightforward to show that the choice of distribution $\pnew(\phi)$ has no influence on $\ptmeld$, it is only a convenient way to define the required transformations.
If $\palt(\phi_\ralt)$ is an alternative distribution with cumulative distribution function $F_\ralt$ we define the transformation
$\phi = \transfun_{\rnew,\ralt}(\phi_\ralt) = F_\rnew^{-}(F_\ralt(\phi_\ralt))$ which preserves marginals on $\phi$ and $\phi_\ralt$.
When we define $\transfun_{m,\ralt}(\phi_\ralt) = F_m^{-}(F_\ralt(\phi_\ralt)) $ we also have $\transfun_{m,\ralt}(\phi_\ralt) = \transfun_{m,\rnew}(\transfun_{\rnew,\ralt}(\phi_\ralt))$ and so
 \[ p_m(\psi_m,Y_m \mid \transfun_{m,\rnew}(\phi)) = p_m(\psi_m,Y_m \mid \transfun_{m,\ralt}(\phi_\ralt)) \]
Similar to~(\ref{e:transform_m}) for a transformation of $\phi_\ralt$ to $\phi = \transfun_{\rnew,\ralt}(\phi_\ralt)$ we have $p_\ralt(\phi_\ralt) = \pnew(\phi)$ and~(\ref{e:copula-melding}) becomes
\begin{equation*}\begin{split}
\ptmeld(\phi,\psi_1,\ldots,\psi_M,Y_1,\ldots,Y_M)
 &= \palt(\phi_\ralt) \prod_{m=1}^M p_m(\psi_m,Y_m \mid \transfun_{m,\ralt}(\phi_\ralt))\\
\end{split}\end{equation*}

The influence of other submodels on submodel~$m$ in the joint model $\ptmeld$ can now be made explicit easily by setting $\palt=p_m$
\begin{equation}\begin{split}
\label{e:copula-influence}
\ptmeld(\phi,\psi_1,&\ldots,\psi_M,Y_1,\ldots,Y_M) \\
 &= p_m(\phi_m) \, p_m(\psi_m,Y_m\mid \phi_m)
\prod_{\ell\neq m} p_\ell(\psi_\ell,Y_\ell \mid \transfun_{\ell,m}(\phi_m))\\
 &= p_m(\phi_m,\psi_m,Y_m) \prod_{\ell\neq m} p_\ell(\psi_\ell,Y_\ell \mid \transfun_{\ell,m}(\phi_m))\\
\end{split}\end{equation}
 with $\phi_\ell=\transfun_{\ell,m}(\phi_m)=F_\ell^{-}(F_m(\phi_m))$ a transformation that preserves the marginals $p_m(\phi_m)$ and $p_\ell(\phi_\ell)$.

Equation~(\ref{e:copula-influence}) shows that the distribution of $\phi_m$ in submodel~$m$ is influenced only through the likelihoods of the transformed variable $\phi_\ell=\transfun_{\ell,m}(\phi_m)$ in the other submodels.
The transformations relate the $\phi_\ell$ so that the quantiles of the distributions $p_\ell(\phi_\ell)$ of all submodels~$\ell$ match.

This form of melding is useful when it is assumed that the marginals of the joint variable are specified correctly for each submodel and are essentially the same, but the variable is expressed on a different scale in each submodel and transformations to a common scale are needed to reveal the common underlying marginal.
Also observe that, although changing $\pnew$ has no influence on $\ptmeld$, for computational reasons it should be chosen so that transformations $\transfun_{m,\rnew}$ can be estimated easily.

\end{document}